\documentclass[twocolumn,trackchanges]{aastex701}

\usepackage{float} 

\newcommand{\NaID}{\ion{Na}{1} D }
\usepackage{tikz}
\usetikzlibrary{arrows.meta, positioning}

\begin{document}

\title{Galactic Rain: Cool Gas Inflows in Red Geyser Galaxies and Their Connection to AGN Activity and Interactions}

\author[orcid=0009-0003-4285-8072]{Arian Moghni}
\affiliation{Department of Astronomy \& Astrophysics, University of California, Santa Cruz, 1156 High Street, CA 95064, USA}
\email[show]{moghni.arian@gmail.com}

\author[orcid=0000-0002-4430-8846]{Namrata Roy} 
\affiliation{School of Earth and Space Exploration, Arizona State University, Tempe, AZ 85281, USA}
\affiliation{Department of Physics \& Astronomy, Johns Hopkins University, Baltimore, MD 21218, USA}
\email{namratar@asu.edu}

\author[orcid=0000-0001-6670-6370]{Timothy M. Heckman} 
\affiliation{Department of Physics \& Astronomy, Johns Hopkins University, Baltimore, MD 21218, USA}
\email{theckma1@jhu.edu}

\author[orcid=0000-0001-9742-3138]{Kevin Bundy}
\affiliation{Department of Astronomy \& Astrophysics, University of California, Santa Cruz, 1156 High Street, CA 95064, USA}
\email{kbundy@ucsc.edu}  

\author[orcid=0000-0003-1809-6920]{Kyle B. Westfall}
\affiliation{University of California Observatories, University of California, Santa Cruz, 1156 High Street, Santa Cruz, CA 95064, USA}
\email{westfall@ucolick.org}  

\author[orcid=0000-0001-6248-1864]{Kate H. R. Rubin}
\affiliation{Department of Astronomy, San Diego State University, San Diego, CA 92182, USA}
\affiliation{Center for Astrophysics and Space Sciences, University of California, San Diego, La Jolla, CA 92093, USA}
\email{krubin@sdsu.edu}

\begin{abstract}
Red geysers are a population of massive ($\log M/M_\odot \sim 10.5$), quiescent galaxies with large-scale but weak, bi-symmetric ionized gas outflows, interpreted as signatures of ongoing, low-level active galactic nucleus (AGN) feedback. We investigate the kinematics and prevalence of cool ($T \sim 100$--$1000$ K), neutral gas traced by \NaID absorption, and its connection to galaxy environment and AGN activity. Using 140 red geyser galaxies from the Sloan Digital Sky Survey-IV Mapping Nearby Galaxies at Apache Point Observatory (MaNGA), we measure spatially resolved velocities and dispersions via double-Gaussian fits to the \NaID doublet. We find that $\sim70\%$ of the cool gas is inflowing, with a median velocity of $\sim47\ \mathrm{km\ s^{-1}}$ ($\sim10\%$ of the expected free-fall velocity), and also exhibits kinematically ordered motions with $\sigma_{\rm NaD}/\sigma_* \sim 0.4$. Additionally, the \NaID absorption is more prevalent in red geysers than in a matched control sample, showing a higher detection fraction (63\% vs 40\%) and reservoir areas $\sim1.6$ times larger. Acceleration ($\sim1$ Myr) and accretion ($\sim20$ Myr) timescales indicate that the Na I D-absorbing clouds are likely young and short-lived. Another intriguing result is that radio-detected red geysers ($30\%$ of the sample) show inflowing gas reservoirs $\sim7$ times larger than in non-radio systems. Similarly, galaxies subject to environmental effects host inflowing gas reservoirs $\sim2.7$ times larger than isolated red geysers. We take this as evidence that galaxy environments play a key role in replenishing the cool gas reservoirs of red geysers, fueling central AGN activity, sustaining radio emission, and regulating long-term quiescence. These findings reveal that quiescent systems are governed by cycles of inflow, feedback, and regulation.

\end{abstract}


\section{\textbf{Introduction}}
One of the central puzzles in modern astronomy is understanding why galaxies stop forming stars and how they remain quiescent over time. Although these ``dead" galaxies could host enough gas to produce new stars, this material does not easily cool and condense to produce new stars \citep[e.g.,][]{2019SSRv..215....5W}. There have been various mechanisms proposed to explain the suppression of star formation, like the feedback from active galactic nuclei (AGN) injecting energy into the surrounding gases in the interstellar medium (ISM), keeping them heated or turbulent and making it more challenging for them to collapse and produce stars \citep[e.g.,][]{2012ARA&A..50..455F}. Despite these advances, it has been challenging to directly observe how these processes operate in quiescent galaxies. 

In recent years, a new class of red and dead galaxies, known as red geysers, has been discovered using the Mapping Nearby Galaxies at Apache Point Observatory (MaNGA) survey \citep{2016Natur.533..504C}. Red geysers make up $\approx6$--$8\%$ of the population of the local quiescent galaxies observed by MaNGA and are characterized by bi-symmetric features in their spatially resolved maps of ionized gas that extend over $\sim 10$ kpc. These ionized gases have unusually high velocities that are misaligned from stellar rotation \citep{2021ApJ...913...33R}

Detailed analyses on the kinematics of these ionized gases have shown that these structures are best explained as bi-conical winds driven by the central AGN \citep{2021ApJ...913...33R, 2021ApJ...922..230R}.

In radio wavelengths, red geysers show evidence for low-luminosity AGN activity. \citealt{Roy_2018} and \citealt{2021ApJ...922..230R} previously used multiple radio surveys, including the Low Frequency Array (LOFAR), the Very Large Array Sky Survey (VLASS), and the Faint Images of the Radio Sky at Twenty-Centimeters (FIRST) to show that a larger fraction of red geysers are detected in radio compared to a matched sample of quiescent galaxies. The detected radio luminosities of red geysers is typically of order $L_{1.4 \text{GHz}} \sim 10^{21}$--$10^{23}$ W Hz$^{-1}$, which is consistent with low-luminosity AGN activity, compared to powerful radio galaxies. They have also shown that the sources of these radio emissions are typically concentrated in the central regions of their host galaxies. In terms of energy, they have been found to be capable of driving mechanical feedback.

The properties of red geysers in radio wavelengths and their galaxy-scale, bi-symmetric ionized gases support the interpretation that these red geysers host low-luminosity AGN that drive large-scale outflows and inject energy into the surrounding interstellar medium.

While the outflows in red geysers have been well studied, it is still unclear how the AGN is fueled and sustained. If the AGN is responsible for maintaining the observed quiescence in these red and dead galaxies, identifying and understanding the fueling process is crucial for completing this picture. A natural candidate for this fuel supply is the cool, neutral gas, which is likely dynamically cold and could thus more easily get funneled in toward the center of the galaxy, where the AGN resides. Additionally, cool gases have previously been discovered in red geysers (\citet{2021ApJ...919..145R}), and if they are not being easily converted into stars, they could instead be participating as a fuel supply for the central black hole.

Observations of star-forming galaxies overwhelmingly find cool neutral gases outflowing across large samples of galaxies with different masses and redshifts (e.g., \citealt{2005ApJ...621..227M}, \citealt{2019SSRv..215....5W}, and \citealt{2005ApJS..160..115R}). Nevertheless, inflowing cool gas has been directly observed in a growing number of galaxies, and the findings suggest that the balance between these inflowing and outflowing gases may depend on a galaxy's evolutionary state (e.g., \citealt{2010ApJ...708.1145K}, \citealt{2025arXiv251011455B}, \citealt{2012ApJ...747L..26R}, and \citealt{2005ApJ...621..227M}).

\citet{2021ApJ...919..145R} provided the first evidence for inflowing cool neutral gas in red geysers using the sodium doublet (\NaID $\lambda\lambda5891, 5897$) absorption, showing that these galaxies contain reservoirs of neutral material despite their quiescence. Their work demonstrated that this gas is kinematically distinct from the ionized gas outflows and mostly inflowing, possibly feeding the central low-luminosity AGNs detected previously in the red geyser sample. 

Following up on these findings, our analysis is motivated by two main questions. First, are the cool, neutral gases being used as a fuel supply for the AGN activity in the center? Second, what is the origin of this gas? Specifically, could this cool gas be externally accreted through galactic interactions with nearby companions as opposed to internal cooling of the warm gases present?

 In this work, we measure the detailed kinematics (velocity and linewidth) of the cool, neutral gas via \NaID absorption in the red geysers sample and examine potential links between the presence of cool gas, the radio activity of the host galaxy, and signatures of interactions with nearby companions. Connecting inflows to radio AGN and environmental signatures allows us to explore how external accretion and feedback work together to shape the life cycle of such quiescent galaxies.

The structure of this paper is as follows. In \S\,\ref{sec:Data and Sample Selection}, we describe the MaNGA Survey, explain the defining features and characteristics of red geyser galaxies, and provide details regarding our classifications of radio detection and interaction status. In \S\,\ref{sec:methods}, we outline our methodology for stellar continuum removal, Gaussian fittings, and error estimations. In \S\,\ref{sec:results}, we present our results on cool gas inflows and their correlations with radio activity and the environment of each red geyser galaxy. In \S\,\ref{section:discussion}, we provide further calculations for the properties of these gases and discuss the implications of our findings for AGN fueling and feedback in quiescent galaxies.

\section{\textbf{Data and Sample Selection}} \label{sec:Data and Sample Selection}
\subsection{MaNGA Survey}
The data for our analysis comes from galaxies observed through the Sloan Digital Sky Survey IV (SDSS-IV) Mapping Nearby Galaxies at Apache Point Observatory (MaNGA; 
\citet{Albareti_2017}; \citet{2017AJ....154...28B}, \citet{2015ApJ...798....7B}; \citet{2015AJ....149...77D}; \citet{2015AJ....150...19L}; \citet{Yan_2016}). MaNGA provides integral field spectroscopy for 10,010 nearby ($z\sim0.03$) galaxies using the SDSS 2.5m telescope \citep{Gunn_2006} and the BOSS spectrographs \citep{Smee_2013}. The spectrographs cover from near-UV wavelengths ($3600\text{\AA}$) to near-IR ($10,000\text{\AA}$), with resolution ranging from $R\approx1400$ at $4000\text{\AA}$ to $R\approx2600$ at $9000\text{\AA}$. The effective spatial resolution is $\sim 2.5''$ (FWHM).

The MaNGA datacubes come in square spatial pixels (spaxels) with a size of $0.5''\times0.5''$, each containing a full spectrum across the wavelength range. These spaxels provide spatially resolved spectral information across each galaxy and allow detailed study of internal processes across the galaxy. For red geyser galaxies, these spaxels typically cover an area of $\sim 0.1\text{ kpc}^2$. 

The galaxies in MaNGA are observed to either 1.5$R_e$ (two-thirds of the sample) or 2.5 $R_e$ (one-third), allowing the coverage of both the inner and outer parts of the host galaxies.

\subsection{Red Geyser Definition and Sample}
Our analysis focuses on red geyser galaxies, a population (140 in total) of recently identified quiescent systems that make up roughly $6$--$8\%$ of the local quiescent population in the MaNGA survey, visually selected based on their characteristic features as described in \citet{2016Natur.533..504C}. The sample and its properties are described in detail in \citet{Roy_2018} and \citet{2021ApJ...919..145R}. 

\begin{figure}[h]
    \centering
    \includegraphics[width=0.45\textwidth]{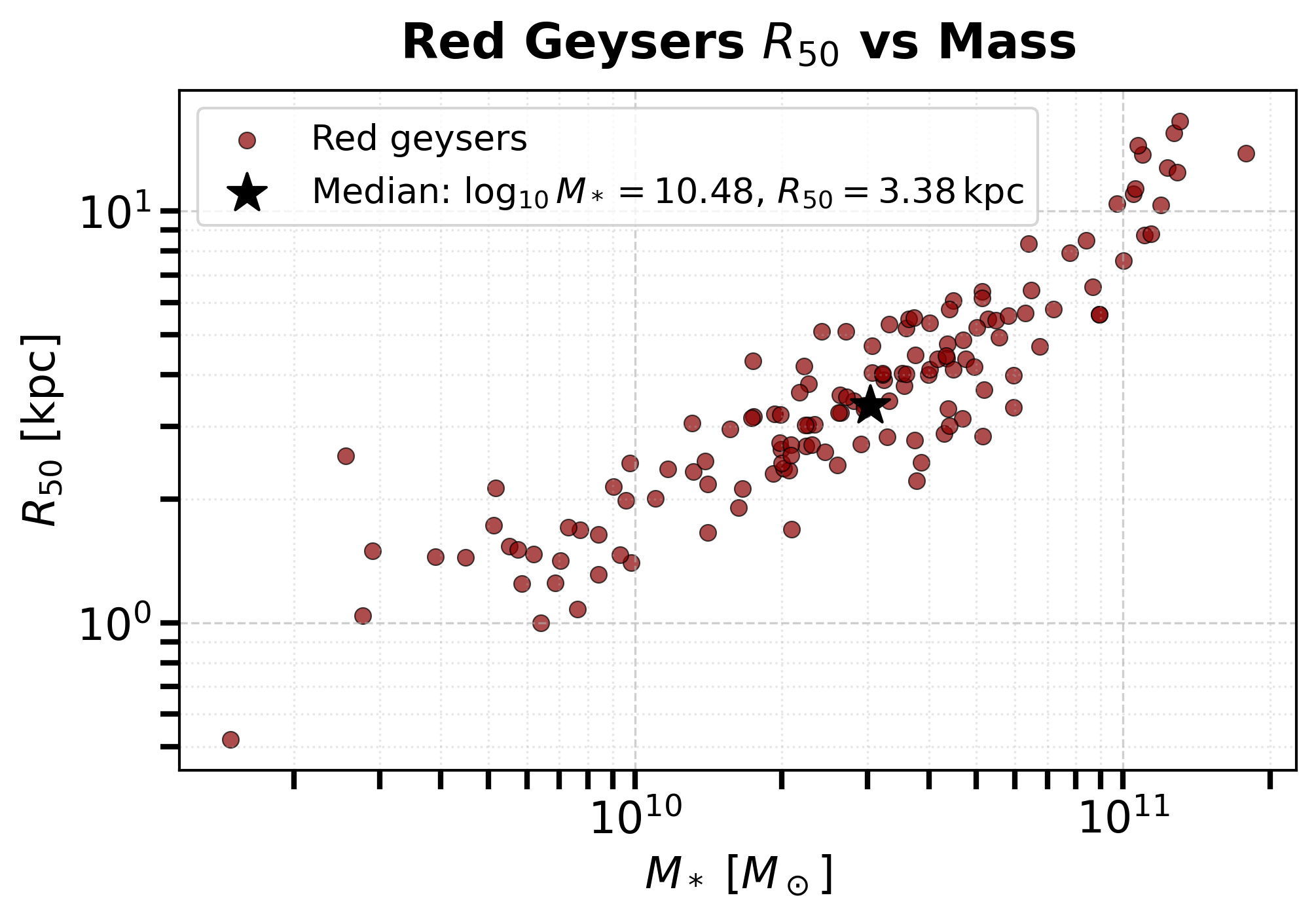} 
    \caption{Half-light radius $R_{50}$ vs stellar mass for the full sample of red geyser galaxies (140 in total). The black star plotted toward the middle represents a characteristic red geyser with mass $\log_{10}M_*/M_\odot =10.48$ and $R_{50}=3.38\text{ kpc}$.}
    \label{fig:radius vs mass}
\end{figure}

Morphologically, red geysers are massive elliptical galaxies without a visible disk or spiral structure. They are effectively quenched, forming stars at incredibly low rates ($\sim 0.01\,M_\odot$\,yr$^{-1}$), yet they exhibit widespread ionized gas with masses of order $10^5-10^6 M_{\odot}$ \citep{2016Natur.533..504C}. Figure \ref{fig:radius vs mass} summarizes our full sample of red geysers, specifically their stellar masses and effective radii $R_{50}$. The median values are $\log_{10}(M_*/M_\odot) = 10.48$ and $R_{50} = 3.38$ kpc.

\begin{figure*}[t]
    \centering
    \includegraphics[width=0.65\textwidth]{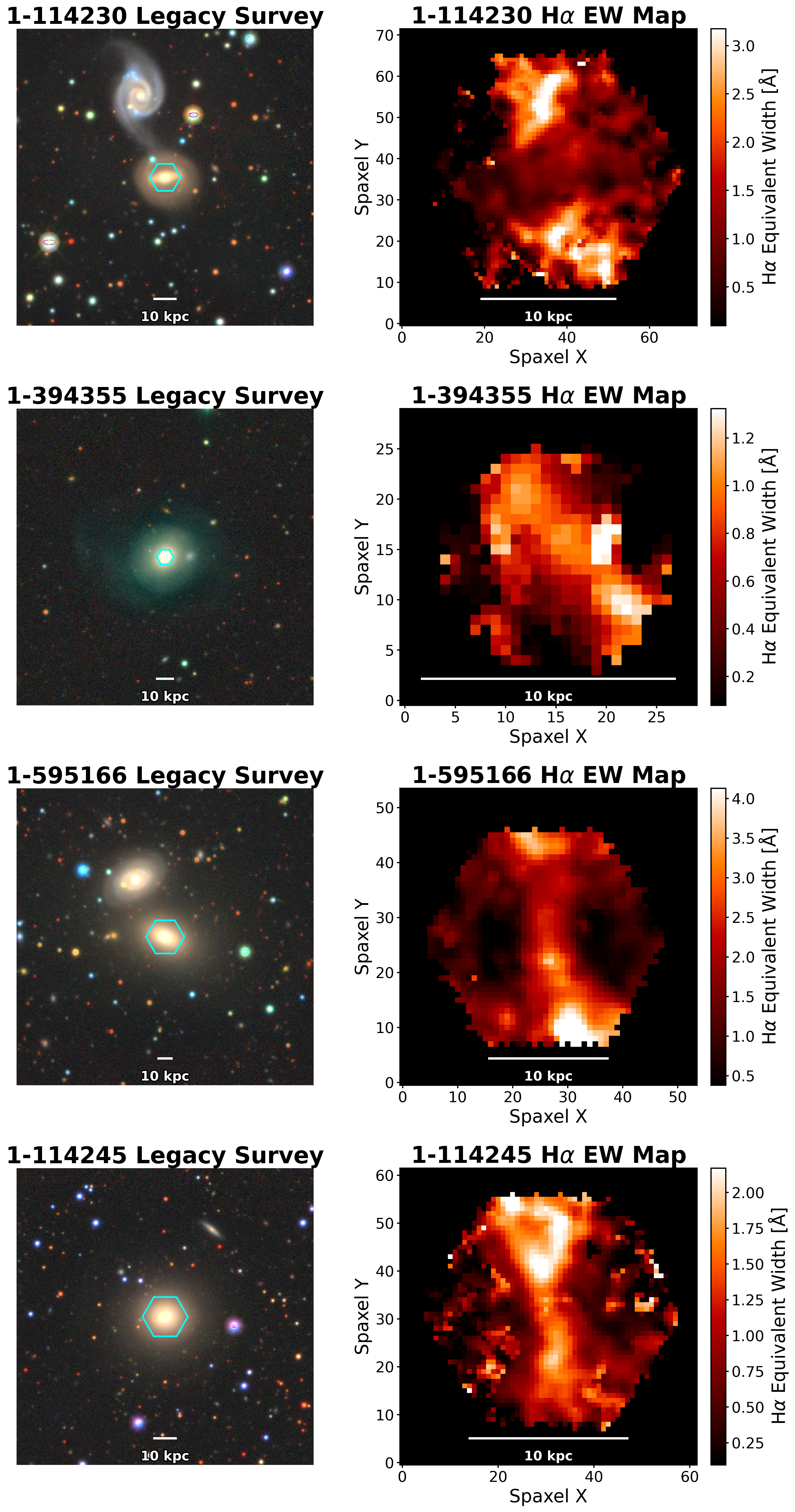} 
    \caption{DESI Legacy Survey images (left) and zoomed-in MaNGA H$\alpha$ equivalent width (EW) maps (right) for four red geyser galaxies (MaNGA IDs: 1-114230, 1-394355, 1-595166, 1-114245). Each Legacy Survey cutout spans $\sim3.5'\times3.5'$ and shows the large-scale morphology and environment. From top to bottom, the panels display red geysers classified as: (1) clear interaction, (2) disturbed without a companion, (3) undisturbed with a companion, and (4) isolated. The H$\alpha$ EW maps (spanning $\sim15''$--$30''$) focus on the MaNGA footprint, outlined by the cyan hexagon, and highlight ionized gas and the characteristic bi-symmetric jet-like features that define red geysers.}

    \label{fig:DESI images and Ha EW}
\end{figure*}

Red geysers are observationally known for their presence of bi-symmetric features in the spatially resolved equivalent width (EW) maps of strong emission lines such as H$\alpha$, [N\,II], and [O\,III]$\lambda$5007 (Figure~\ref{fig:DESI images and Ha EW}). These are galaxy-scale ionized outflows extending over $\sim10\text{ kpc}$ and are accompanied by unusually high gas velocities, reaching up to $\pm300$\,km\,s$^{-1}$. Another feature of the ionized gases in red geysers is that they have large velocity dispersions, typically between $\sim 220$--$250$\,km\,s$^{-1}$. 

Red geysers are therefore interpreted as active galactic nucleus (AGN) feedback in action (\citealt{2016Natur.533..504C}; \citealt{2021ApJ...913...33R}). It has been confirmed that red geysers host low-luminosity radio AGN activity, which are the possible drivers of their observed large-scale winds (\citealt{Roy_2018}; \citealt{2021ApJ...922..230R}).

Their spatially-resolved emission-line ratios on Baldwin-Phillips-Terlevich (BPT) diagrams \citep{1981PASP...93....5B} place most spaxels in the LINER or Seyfert regimes, suggesting that the ionized gas is largely powered by evolved stellar populations (post-asymptotic giant branch stars) with possible contributions from shocks and AGNs.

\subsection{Radio Detection Classifications}
The radio properties of red geysers have previously been studied by \citet{2021ApJ...922..230R} and \citet{Roy_2018}. With data from multiple surveys across a wide range of frequencies --- the LOFAR Two-Metre Sky Survey (LoTSS; 150 MHz), the Faint Images of the Radio Sky at Twenty-Centimeters (FIRST; 1.4 GHz), and the VLA Sky Survey (VLASS; 3 GHz) --- they found that 42 of the 140 red geysers are radio detected. Out of the $\sim75\%$ resolved in LoTSS or FIRST, they show that the size of the radio sources typically ranges from a few kpc to $\sim 25\text{ kpc},$ with a few extending beyond $40\text{ kpc}.$ For most of the detected systems, the radio sources have compact morphologies and are consistent with low-power radio AGN.

The sources that are extended tend to be found in red geysers with the lowest specific star formation rate. This could be due to the stronger effects of the AGN on the host interstellar medium, or simply more massive halos.

In our paper, following the classification established by \citet{2021ApJ...922..230R}, we classify 42/140 of our red geysers as radio-detected. The rest, 98/140, are labeled as non-radio detected. These groupings allow us to better analyze the connections between the cool, neutral gases and the AGN activity as we discuss in $\S$\ref{sec:methods}.

\subsection{Interaction Classifications}
An important part of our analysis is understanding how the environments of red geyser galaxies influence their evolution. 

We use the DESI Legacy Imaging Surveys \citep[][example in Fig. \ref{fig:DESI images and Ha EW}]{2019AJ....157..168D} to visually classify our sample of 140 red geyser galaxies as one of four categories:

\begin{enumerate}
    \item \textbf{Clear interaction}: systems showing clear tidal tails or merging signatures. 
    
    \item \textbf{Disturbed without companion}: galaxies that appear irregular and asymmetric but lack a clear nearby  neighbor. These systems likely correspond to past/minor mergers.
        
    \item \textbf{Undisturbed with companion}: galaxies that appear morphologically undisturbed but still have a close ($\lesssim150 \text{ kpc}$) neighbor.  
    
    \item \textbf{Isolated}: systems that appear undisturbed with no clear sign of a companion.
    
\end{enumerate}

To take into account possible contamination from background galaxies appearing as ``nearby neighbors," we check the redshifts provided by the DESI Legacy Imaging Surveys for the nearby companion candidates. In each case, we compare the companion's redshift to the corresponding red geyser and ensure that the difference in redshift remains within $\Delta z\sim 0.001$. Using this threshold, which corresponds to a velocity difference of $\sim 300 \text{ km s}^{-1}$, we remove contamination from background galaxies and ensure robust analysis of the environments of red geyser galaxies.

We find that out of 140 red geysers, 29 are clearly interacting, 18 are disturbed without a companion, 49 appear undisturbed with a companion, and 44 are isolated. A summary of our classifications is shown in Table \ref{table:classifications}. 

\begin{table}[h!]
\centering

\caption{Interaction Class vs. Radio Detection.\\ ``R'' = Radio Detected; ``NR'' = Non-Radio Detected. Radio-detection status is adopted from \citet{2021ApJ...922..230R}, while the interaction classes are identified with visual inspections through the data from the DESI Legacy Imaging Surveys \citep{2019AJ....157..168D}.}
\label{table:classifications}
\begin{tabular}{|l|c|c|}
\hline
\textbf{Interaction Class} & \textbf{R (42)} & \textbf{NR (98)} \\
\hline
Clear interaction (29)        & 12 & 17 \\
\hline
Disturbed, no companion (18)  & 5  & 13 \\
\hline
Undisturbed, close companion (49) & 19 & 30 \\
\hline
Isolated \& symmetric (44)    & 6  & 38 \\
\hline

\end{tabular}
\end{table}

For our analysis of \NaID across the different interaction groups, we combine the first three categories --- clear interactions, disturbed systems without companions, and undisturbed systems with nearby companions --- into a single class of \textit{environmentally influenced} galaxies. Despite their morphological differences, these systems all represent galaxies that are not isolated and may therefore be influenced by environmental processes, such as ongoing or past mergers as well as the presence of nearby neighbors.

This grouping not only allows us to better understand the role of the environments of red geysers in their evolution but also helps us maintain sufficient sample sizes for robust statistical comparisons. Overall, there are 96 \textit{environmentally influenced} red geysers and 44 \textit{isolated} systems.

\section{\textbf{Methods}}
\label{sec:methods}

\subsection{Tracing Cool, Neutral Gas with the \NaID Doublet}

To study the cool, neutral gas in red geyser galaxies, we use the sodium doublet (\NaID $\lambda\lambda5891, 5897\text{\AA}$). These absorption lines come from the transition $3s\rightarrow3p$ in the neutral sodium atom, with the doublet arising from spin-orbit coupling, which is a quantum mechanical effect where the orbital angular momentum of an electron interacts with its own intrinsic spin. Since \NaID is a transition from the ground state, it can produce strong absorption without requiring significant column densities, making it a reliable tracer of cool gases in the interstellar medium of galaxies (\citealt{1992MNRAS.259...47C}, \citealt{2012A&A...545A..21P}).

Another useful property of sodium is that it has an ionization potential of only 5.14 eV, which is much lower than the 13.6 eV required to ionize neutral hydrogen. As a result, sodium is easily ionized, and its absorption requires gas that is predominantly in the neutral phase.

Because of these properties, \NaID has been widely used not only as a tracer of the cool, neutral gas in the interstellar medium of galaxies, but also as a tracer of cool, neutral gas participating in large-scale winds (e.g., \citealt{2000ApJS..129..493H}, \citealt{2022ApJ...936..171R}, and \citealt{2024MNRAS.528.4976D}).

For our study of red geyser galaxies, we use \NaID to better understand the role of this cool gas in possibly fueling and sustaining the AGN activity in the center and its connection to the large-scale environment of the host galaxy.

\subsection{Removing Stellar Absorption}


To study the cool, neutral gas in the interstellar medium traced by the \NaID doublet, it is essential to first remove the stellar component of the absorption lines. This is especially important in the case of red geyser galaxies whose evolved stellar populations (dominated by K--M type stars) have deep \NaID features \citep{1984ApJS...56..257J, 1991ApJ...367..547T} that typically contribute much more to the observed feature than does absorption by interstellar gas.  
Here, we use a full spectral fitting approach to model each galaxy spectrum as a superposition of template spectra that represent the stellar component and the nebular emission.  Only the stellar component contributes directly to the \NaID region in the model, and we mask the observed \NaID  region during the fit.  This allows us to use the model best fit to the remainder of the spectrum to predict the \NaID absorption that is only due to stellar atmospheres, which we then remove for the remainder of our analysis. This section describes our procedure in detail.

We use the MaNGA Data Analysis Pipeline (DAP) to model the spectra of our red geyser sample.  The DAP is the software package developed to provide high-level quantities distributed by the MaNGA Survey, such as stellar kinematics and nebular emission-line properties.  The survey-level execution of the DAP for these products is described in detail by \citet{2019AJ....158..231W} and \citet{2019AJ....158..160B}, with updates relevant to the final MaNGA data release (SDSS DR17) presented by \citet{2022ApJS..259...35A}.  The DAP is still maintained, with recent updates and software releases described in its online documentation.\footnote{\url{https://sdss-mangadap.readthedocs.io}}  We take advantage of the built-in flexibility of the DAP to re-analyze our subsample of MaNGA datacubes, with the primary difference being the mask applied to the \NaID region.


Briefly, the DAP modules relevant to our analysis spatially bin the spectra to meet a minimum S/N threshold, measure the stellar kinematics, and simultaneously model the stellar continuum and nebular emission lines.  The workflow is as follows.

First, the DAP selects spaxels for analysis based on two criteria: 
\begin{itemize}
    \item Spectral coverage: At least 80\% of the pixels in each spectrum must not be masked, and
    \item S/N ratio: The $g$-band-weighted S/N must be at least 1.
\end{itemize}
For spaxels that satisfy these criteria, the DAP groups them into Voronoi bins \citep{2003MNRAS.342..345C} until they reach an average S/N of 10.  Importantly, not all spectra will meet a S/N threshold of 10 even after the binning procedure is complete because of the limited field-of-view of the observations.  Given this, we limit all remaining analysis to only those spaxels or bins with S/N$\geq 10$.

Second, the DAP measures the stellar kinematics using pPXF \citep{2023MNRAS.526.3273C}; spectral regions with nebular emission lines are masked and we use the hierarchically clustered set of MILES \citep{2007yCat..83710703S} templates presented by \citet{2019AJ....158..231W}.  We fit the first two moments ($V$ and $\sigma$) of the stellar line-of-sight velocity distribution.  Importantly, we note that the MILES templates do not cover the full spectral range of MaNGA, only spanning $\sim$3600 -- 7400${\rm \AA}$.

Finally, the DAP uses the results from this fit to fix the stellar kinematics while refitting the galaxy spectra with a mix of stellar-continuum templates and nebular emission lines.  For the survey-level execution of the DAP, the primary purpose of this module was to measure the emission-line properties.  For our purposes, however, we use it to provide the best-fit continuum normalization that removes the stellar component of the \NaID feature.  

The stellar continuum templates we use in this module are stacks of stellar spectra with similar features based on
hierarchical clustering of MaStar spectra \citep{2019ApJ...883..175Y} provided by \citet{2020MNRAS.496.2962M}. To assess the sensitivity of our results to the choice of stellar templates, we repeated the full Na I D fitting procedure described in \S\,\ref{subsection: gaussian fitting} for a representative galaxy using an alternative template with simple stellar-population (SSP) models. We find that the calculated \NaID velocities are highly robust, differing by less than $1\%$ (median ratio $\sim0.996$), with velocity maps that are visually indistinguishable between the two template choices. The velocity dispersions are also stable, showing similar features in their maps and a modest systematic offset of $\sim 8\%.$ Similarly, the maps of \NaID equivalent width remain visually indistinguishable, but their values exhibit an offset of order $\sim20\%$. Equivalent width values are not directly used in any of our analysis, and the choice of our templates does not affect any of our quantitative or qualitative conclusions throughout the paper.

\begin{figure*}[t]
    \centering
    \includegraphics[width=0.8\textwidth]{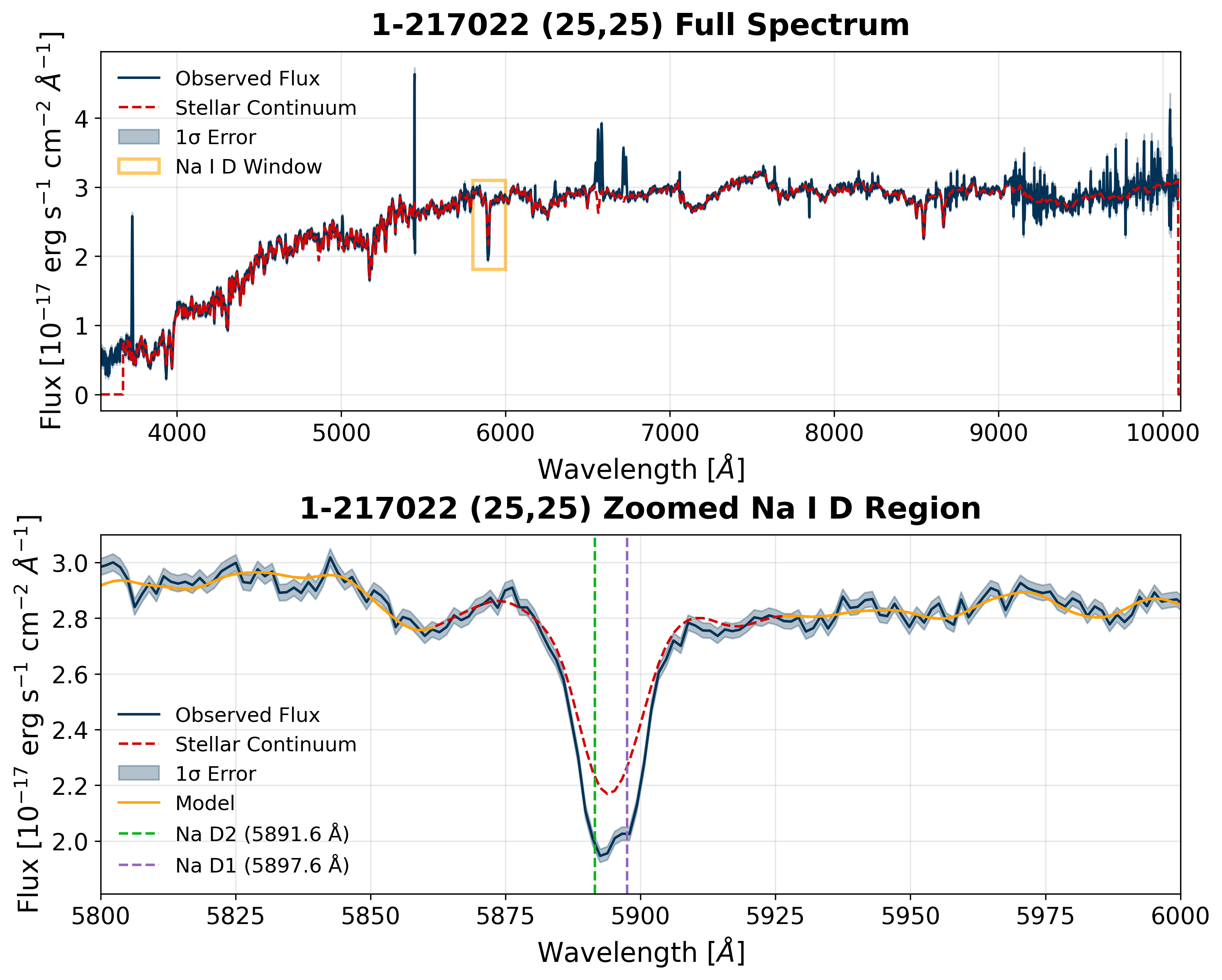} 
    \caption{Example MaNGA spaxel spectrum (spaxel x, y of 25, 25) from the red geyser galaxy 1-217022. Top: Full observed spectrum (blue) with the DAP stellar continuum model (red dashed line) overplotted. The spectrum shows absorption and emission features across the optical range, illustrating the quality of the DAP fit to the stellar continuum. Bottom: Zoom-in around the \NaID doublet, which is indicated in the orange box in the galaxy's spectrum at the top. The observed flux (blue), stellar continuum from DAP (red), and DAP absorption model (yellow; stellar absorption + emission) are shown. The model here excludes the absorption feature, indicating the region over which the fitting was not done. The positions of D2 and D1 lines are shown at $\lambda5891.6\text{\AA}$ (green dashed)  and $\lambda5897.6\text{\AA}$ (purple dashed), respectively. Additionally, $1\sigma$ error of the observed flux is shaded.}
    \label{fig:full_dap_spectrum}
\end{figure*}

Once the continuum fit is obtained (as shown in Figure \ref{fig:full_dap_spectrum}), the observed galaxy spectrum is divided by this stellar model to yield a residual spectrum, where the stellar absorption has been removed, leaving behind only interstellar absorption.

\begin{figure}[h]
    \centering
    \includegraphics[width=0.5\textwidth]{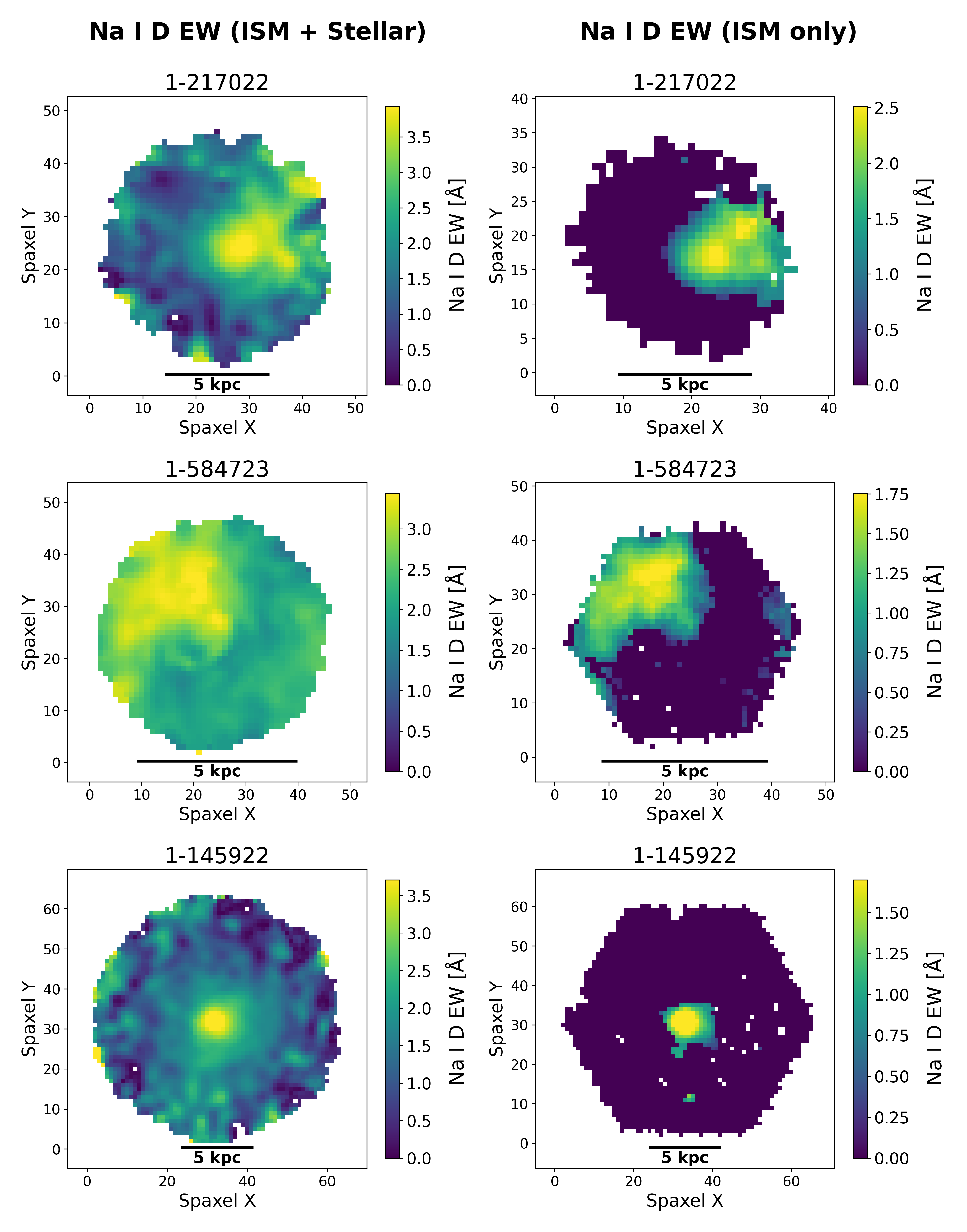} 
    \caption{Spatial maps of \NaID equivalent width (EW; within a window of $5876.88\text{\AA}-5909.38\text{\AA}$) for three red geyser galaxies in three different rows with the corresponding MaNGA ID above each panel. The left column shows the total \NaID EW observed, including the contributions from both the interstellar medium and the stellar atmospheres. The right column shows the \NaID EW after subtracting the stellar component (ISM only). After the $S/N>5$ cut that we describe in \S\,\ref{sec:methods}, the majority of the spaxels with $EW\approx0$ would be removed. These maps highlight that the stellar, absorption dominates the observed \NaID feature in these galaxies.}
    \label{fig:ew_before_after}
\end{figure}

Throughout this paper, it is this residual that we use to analyze the \NaID absorption and its kinematic signatures. The \NaID absorption maps of three red geyser galaxies before and after removing the stellar absorption contributions is shown in Figure \ref{fig:ew_before_after}. Here, it is clear that the stellar absorption removal has a significant effect on the resultant maps.

\subsection{Gaussian Fitting}
\label{subsection: gaussian fitting}
To extract the kinematics of the neutral gas, we fit the residual \NaID absorption profile in each spaxel with a double-Gaussian model. The normalized flux is modeled using the following expression:

\begin{equation}
F(\lambda) = 1 - \left[ G_{\mathrm{D2}}(\lambda) + G_{\mathrm{D1}}(\lambda) \right],
\end{equation}

where each Gaussian component is defined as

\begin{equation}
G(\lambda) = A \, \exp \left[ -\frac{1}{2} 
\left( \frac{\lambda - \left(\lambda_{0} + \Delta\lambda\right)}{\sigma} \right)^{2} \right].
\end{equation}
Here, $A$ is the amplitude, $\sigma$ is the standard deviation (related to the velocity dispersion of the absorbing gas), and $\lambda_{0}$ is the rest-frame wavelength of the transition ($\lambda_{\mathrm{D2}} = 5891.58 \,\text{\AA}$, $\lambda_{\mathrm{D1}} = 5897.56 \,\text{\AA}$). The wavelength shift $\Delta\lambda$ is defined as

\begin{equation}
    \Delta\lambda = \frac{v_{offset}}{c}* \lambda_0
\end{equation}
where $c$ is the speed of light, and $v_{offset}$ is the line-of-sight velocity of the gases relative to the mean galaxy redshift --- both in $\mathrm{km\,s^{-1}}$. Redshifted (positive) gas velocities indicate inflowing motion, while blueshifted (negative) velocities indicate outflow.

The velocity offset $v_{\rm offset}$ and the velocity dispersion $\sigma$ for both Gaussians are set to be equal. Additionally, for our fits, there is a constraint on the amplitudes of the two Gaussians that forces the weaker one at the longer wavelength (D1) to have an amplitude between 0.5 to 0.99 times the amplitude of the Gaussian in the shorter wavelength (D2). This is because in the optically thin conditions, the D2 line absorbs about twice as much light as D1 because the oscillator strength ratio $\frac{f_{D2}}{f_{D1}}=\frac{2}{1},$ so in an optically thin regime, D2 will have twice the line strength and thus twice the absorption depth of D1. However, as we approach the optically thick regime, their ratio approaches 1.

Throughout our fitting in each spaxel, we calculate the signal-to-noise ratio (S/N) within the \NaID passband ($5871.9 \text{\AA}-5914.4\text{\AA}$). The signal is defined as the depth of the residual \NaID absorption feature, while the noise is estimated as the standard deviation of the residual continuum --- excluding the main \NaID absorption region. For our analysis, we cut out any spaxel with a S/N $<$ 5. While the S/N of 10 discussed in \S\,\ref{sec:methods} is used for modeling the spectra of our galaxies and ensuring a less noisy observed spectrum overall, this S/N cut of 5 allows us to remove any spaxels with negligible \NaID absorption from the ISM after we have removed the contributions from the stellar atmospheres.

\begin{figure}[h]
    \centering
    \includegraphics[width=0.45\textwidth]{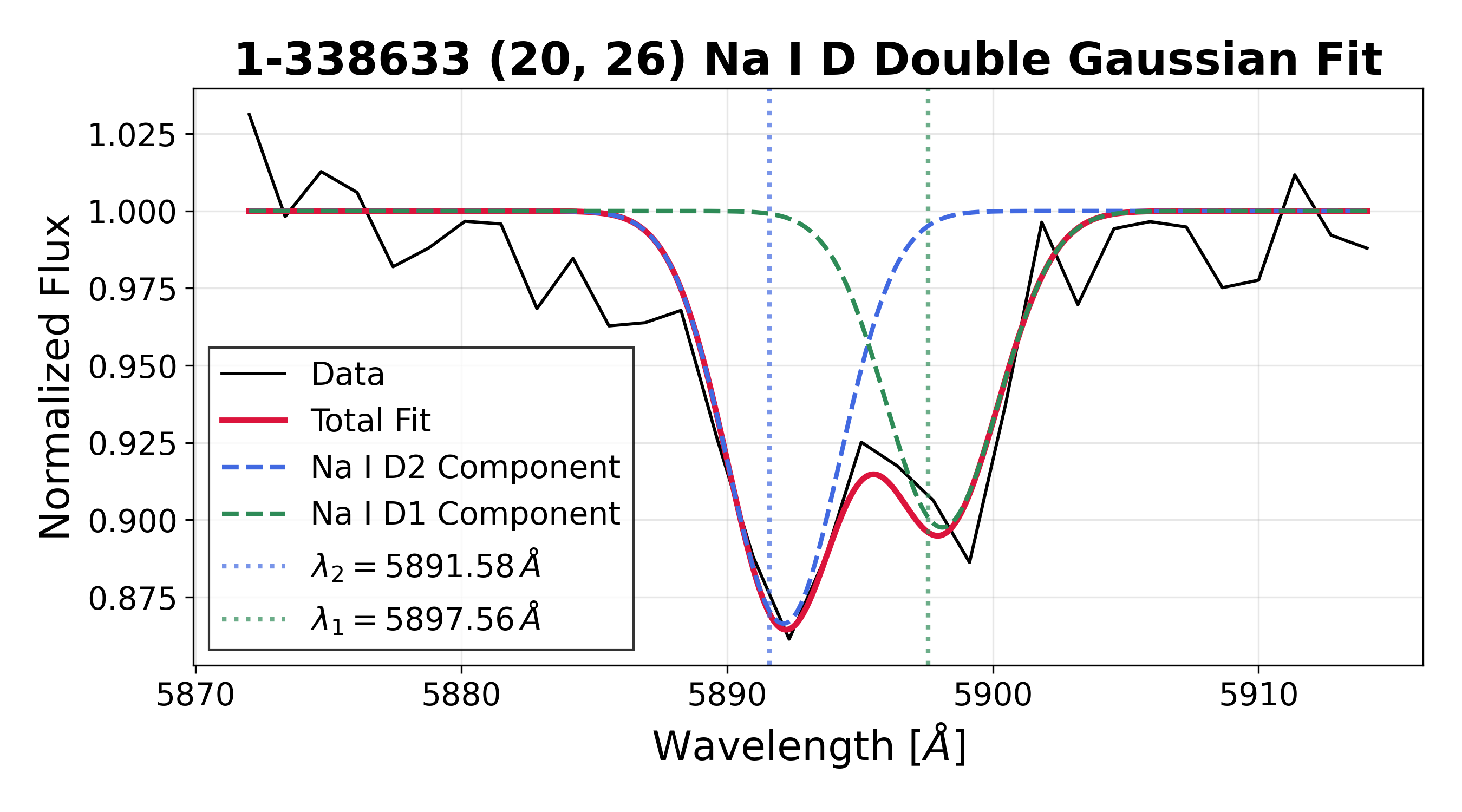} 
    \caption{Example of a double-Gaussian fit to the \NaID absorption profile in a single spaxel (20,26) of the red geyser galaxy 1-338633. The black line shows the residuals after removing the stellar component of the absorption, and the red curve represents the total fitted profile. The individual Gaussian components for the D2 ($\lambda_{0} = 5891.58$ $\text{\AA}$) and D1 ($\lambda_{0} = 5897.56$ $\text{\AA}$) lines are shown as blue and green dashed lines, respectively. Vertical dotted lines show the rest-frame wavelengths of the two transitions. The fits allow us to extract the velocity offset and velocity dispersion of the neutral gas relative to the systemic velocity of the galaxy.}
    \label{fig:double-Gaussian fit}
\end{figure}

From the fitted parameters, we extract both the velocity offset, $v_{offset}$, and the dispersion, $\sigma$, of the absorbing gas. An example of a double-Gaussian fit is shown in Figure \ref{fig:double-Gaussian fit}. Here, the figure demonstrates how the double-Gaussian model accurately reproduces the residual \NaID absorption profile, allowing us to extract the velocity offset and the dispersion of the gas clouds from the fits.
To assess the uncertainties in our double-Gaussian fits, we adopt a Monte Carlo approach. We first estimate the noise level of the continuum by computing the standard deviation, $\sigma$, of the continuum on the blue and red sides of the absorption line (excluding the feature itself). Then, we randomly perturb the data points across the wavelength range over which the double-Gaussian fitting is done by adding to each data point a number between $-\sigma$ and $+\sigma$. This process is repeated 50 times per spaxel, and for each round, the fitting parameters are extracted separately. The standard deviation of those results yields an error for our fits. 

To ensure that our analysis on the kinematics of the cool gases is robust, we remove spaxels that have a $v_{offset}$ uncertainty greater than 50 km/s and a $W50/W50*$ error larger than $40\%$. A visual inspection of these spaxels confirm that detections not passing these criteria generally correspond to poorly constrained \NaID fits.

Across our sample of spaxels in every red geyser galaxy, the median percent error in velocity dispersion is $7.86\%$ while the typical uncertainty in the velocity offset is $13.84 \text{ km s}^{-1}$. These uncertainties arise from the fact that the absorption features that we are fitting are noisy.

Going forward, we use velocity width at half maximum, $W50$, and use that as our measure of line broadening. This allows us to better track asymmetric or broad profiles, and determine the gas dynamics.

An additional important step is to take into account the instrumental broadening because the spectrograph can smear the signal due to its finite resolution, so we have to remove the instrumental contribution. We convert our $W50_{\text{obs}}$ to $W50_{\text{true}}$ using the following equation:
\begin{equation}
    W50_{\text{true}} = \sqrt{W50_{\text{obs}}^2 - W50_{\text{instrument}}^2} ,
\end{equation}
where the $W50_{\text{instrument}}$ for the \NaID line is 165 $\mathrm{km\,s^{-1}}$. This comes from the fact that MaNGA's spectral resolution is $R\approx2000$ for NaD; since $R=\frac{\lambda}{\Delta\lambda}$, where $\lambda$ $\approx 5900\text{\AA}$ and $\Delta\lambda$ is the smallest difference in wavelength that can be resolved at that $\lambda$, re-arranging the equation and converting our units, we get $\Delta\lambda=165\ \mathrm{km\,s^{-1}}$.

With these measurements and corrections in place, we now turn to the results of our analysis, focusing on the kinematic properties of the neutral gas in red geyser galaxies and the properties of the host galaxy.

\section{\textbf{Results}} \label{sec:results}
\subsection{Enhanced Prevalence of \NaID in Red Geysers}

A critical first step for our analysis is to quantify the prevalence of \NaID absorption in the ISM of red geysers and compare that to a matched sample of control galaxies.

To do this comparison, following \citet{Roy_2018}, we first construct a control sample of galaxies (139 in total) that match red geysers in their stellar mass, redshift, rest frame $NUV-r$ color, axis ratio, and presence of ionized gas, ensuring that any observed \NaID differences is not due to basic properties of their host galaxy. We note that the stellar absorption in the control sample is removed using templates based on simple stellar populations. To avoid potential biases arising from differences in stellar templates, we make sure to apply the same templates to the red geysers in this comparison.

We quantify the prevalence of the cool, neutral gas by calculating the summed area of the detected \NaID spaxels in each galaxy, representing the physical size of their gas reservoirs. For systems that we observe to have zero detections of \NaID with $S/N>5$, we place an upper limit to their area by manually adding the equivalent area of one spaxel within the host galaxy and separating them from the systems with detected Na I D. In these separate ``upper limit" bins, we also include the \NaID area in the populations of galaxies with $<5$ detections. This allows us to separate detections and non-detections while still including systems without appreciable Na I D. We repeat these measurements across red geysers and the control sample and create distributions of the area of the \NaID reservoirs for both populations. 

\begin{figure}[h]
    \centering
    \includegraphics[width=0.5\textwidth]{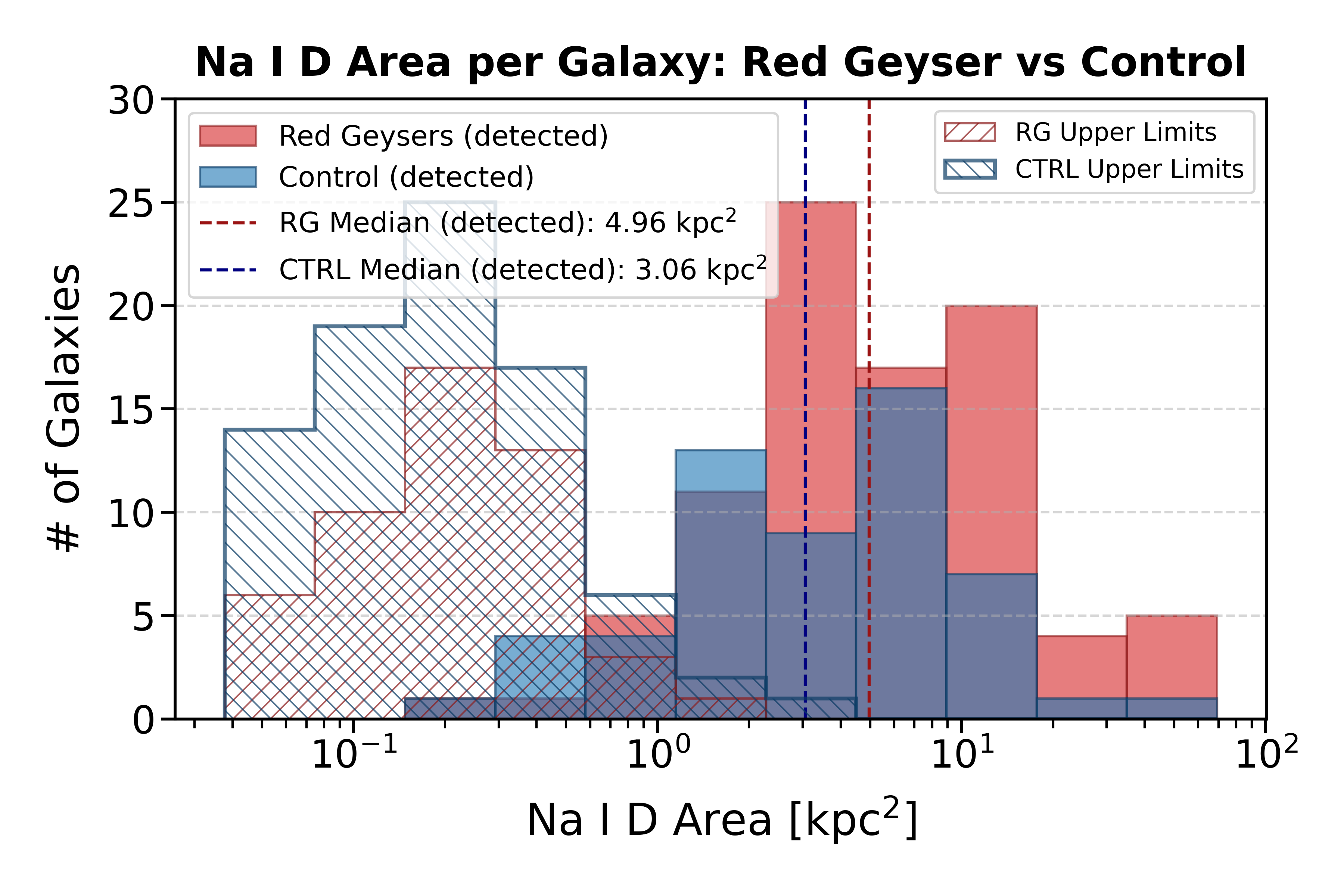} 
    \caption{Distribution of \NaID reservoir area per galaxy for red geysers (red) and the control galaxies (blue). Solid histograms represent galaxies with at least one detected \NaID spaxel, while the hatched histograms indicate upper limits for galaxies with insufficient detections of cool, neutral gas. The vertical dashed lines indicate the median area of \NaID reservoirs in red geysers (red; $4.96 \text{ kpc}^2$) and the control galaxies (blue; $3.06 \text{ kpc}^2$). Red geysers host systematically larger \NaID reservoirs, both in terms of detection fraction ($63\%$ in red geysers vs $40\%$ in control galaxies) and in the total area of \NaID reservoirs in galaxies with detected Na I D.}
    \label{fig:NaD Prevalence Histogram}
\end{figure}

Figure \ref{fig:NaD Prevalence Histogram} shows that the median area of the detected \NaID ($\ge5$ detections) in red geysers is $\sim 1.6$ times larger than the detected cool gas reservoirs of the control sample of galaxies (median $4.96 \text{ kpc}^2$ vs $3.06 \text{ kpc}^2$). Our Mann-Whitney U test on the two detected distributions yields a p-value of 0.025, indicating that the observed difference is statistically significant. 

Another important point is that the detection fraction of red geysers (defined as galaxies with at least 5 detected spaxels) is calculated to be $63\%$, which is larger than the control sample's detection rate of $40\%$. 

The correlations in our findings are in agreement with \citet{2021ApJ...919..145R}, where the detection fractions are measured to be $50\%$ and $25\%$ in red geysers and the control sample, respectively. The modest difference in the number of detections could be due to the signal-to-noise requirements, differences in the stellar continuum subtraction, and errors associated with the double-Gaussian line fits.

Overall, the detected \NaID is more prevalent in red geysers than a matched control sample of galaxies.

\subsection{Redshifted, Narrow Lines}
By repeating the double-Gaussian fits to our \NaID absorption profiles and extracting the velocity and dispersion of the gas clouds in spaxels with a $S/N >5$ in every red geyser galaxy, we then begin our analysis on the kinematics of the cool, neutral gases in these systems.

\begin{figure}[h]
    \centering
    \includegraphics[width=0.5\textwidth]{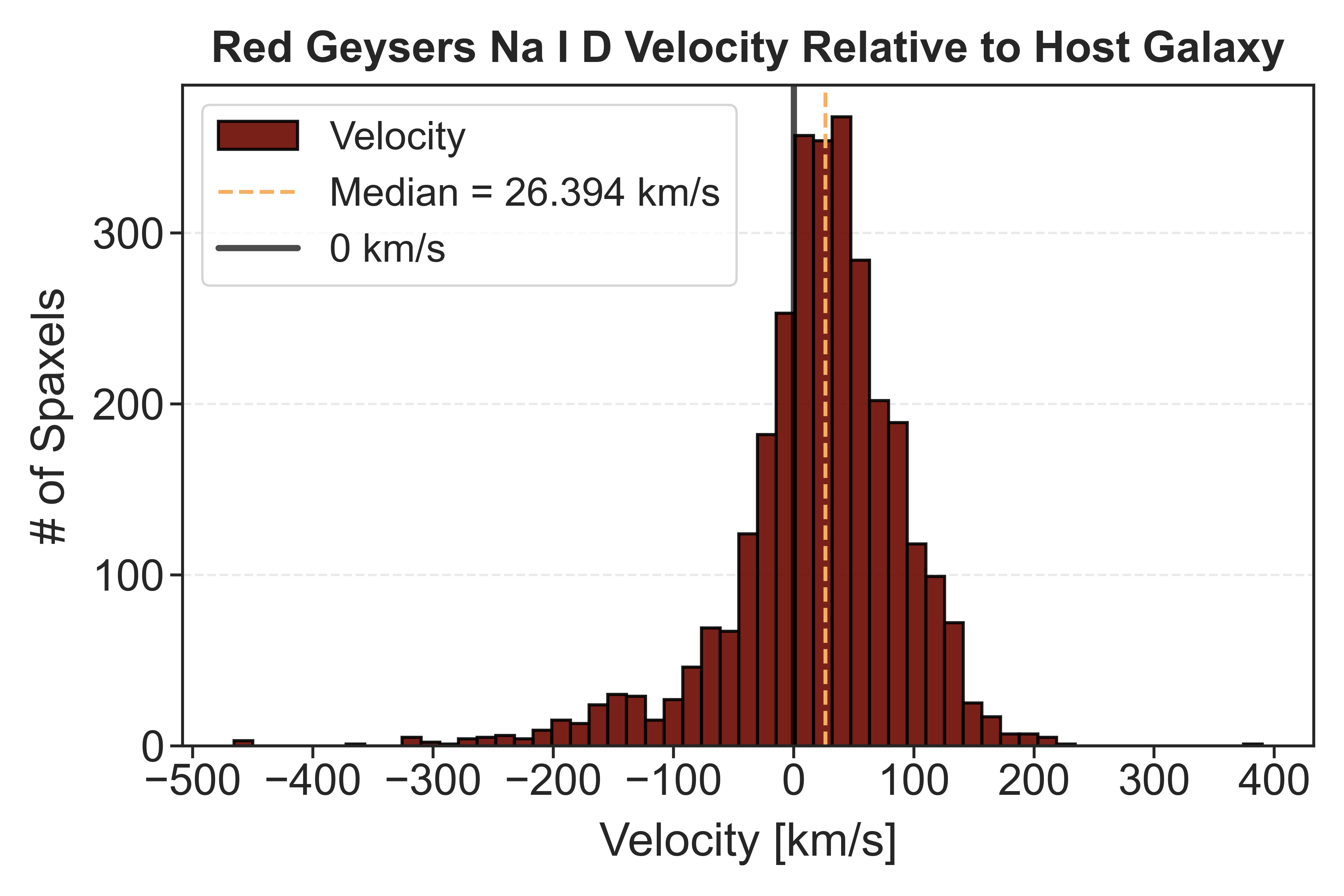} 
    \caption{Distribution of the velocities of the cool gases (\NaID) found through the double-Gaussian fittings of spaxels with $S/N>5$ in every red geyser galaxy. The blue vertical line shows where the velocities are zero. The histogram shows that the detections of inflowing spaxels outnumber the gas clouds with outflowing velocities by a factor of $>2$. Additionally, the median velocity across the inflows and outflows is shown using the orange vertical line at $26.394 \text{ km}\,\text{s}^{-1}.$}
    \label{fig:NaD velocities histogram}
\end{figure}

We first examine the distribution of the \NaID absorption velocities relative to the systemic velocity of the host galaxy. Figure \ref{fig:NaD velocities histogram} shows the histogram of all extracted velocities across the full red geyser sample. Positive velocities (redshifted) indicate that the gas is moving toward the galaxy along our line of sight (inflowing), while negative velocities (blueshifted) correspond to gases that are moving toward the observer away from the galaxy along our line of sight. 

While outflowing gas on the far side of the galaxy can in principle produce redshifted absorption, such gas is silhouetted against only a small fraction of the background starlight and therefore would produce a very weak absorption. On the other hand, gases residing closer to us, in the near side of the galaxy, have more background starlight to absorb and would thus dominate the observed absorption detections. Therefore, we interpret the redshifted and blueshifted \NaID discussed throughout this section as tracing cool gas on the near side of the galaxy, moving toward (inflow) or away from (outflow) the host galaxy along our line of sight.

We observe that across our sample of 140 red geyser galaxies, $\sim 70\%$ of all of the spaxels with appreciable \NaID absorption are inflowing, and these redshifted spaxels have a median velocity of $47\text{ km s}^{-1}$. These inflowing gases support the infalling signatures found in \citet{2021ApJ...919..145R}. 

On the other hand, the spaxels with blueshifted velocities, are less numerous, making up $\sim 30\%$ of all spaxels with \NaID and are found in $\sim 40\%$ of our sample of red geyser galaxies (compared to the $\sim60\%$ for inflowing). These outflowing gases have median velocities of $-34\text{ km s}^{-1}.$ In total, with inflowing and outflowing gases, the median velocity for these gas clouds across the reliable ($S/N>5$) spaxels of every red geyser galaxy is $\sim 26 \text{ km s}^{-1}$.

Additionally, we normalize the $W50_{NaD}$ of each spaxel by its corresponding stellar $W50_*$ found through the DAP fits described in \S\,\ref{sec:methods}. For this normalization, we ensure that both the $W50_{NaD}$ and the $W50_*$ have been corrected for instrument broadening.

\begin{figure}[h]
    \centering
    \includegraphics[width=0.5\textwidth]{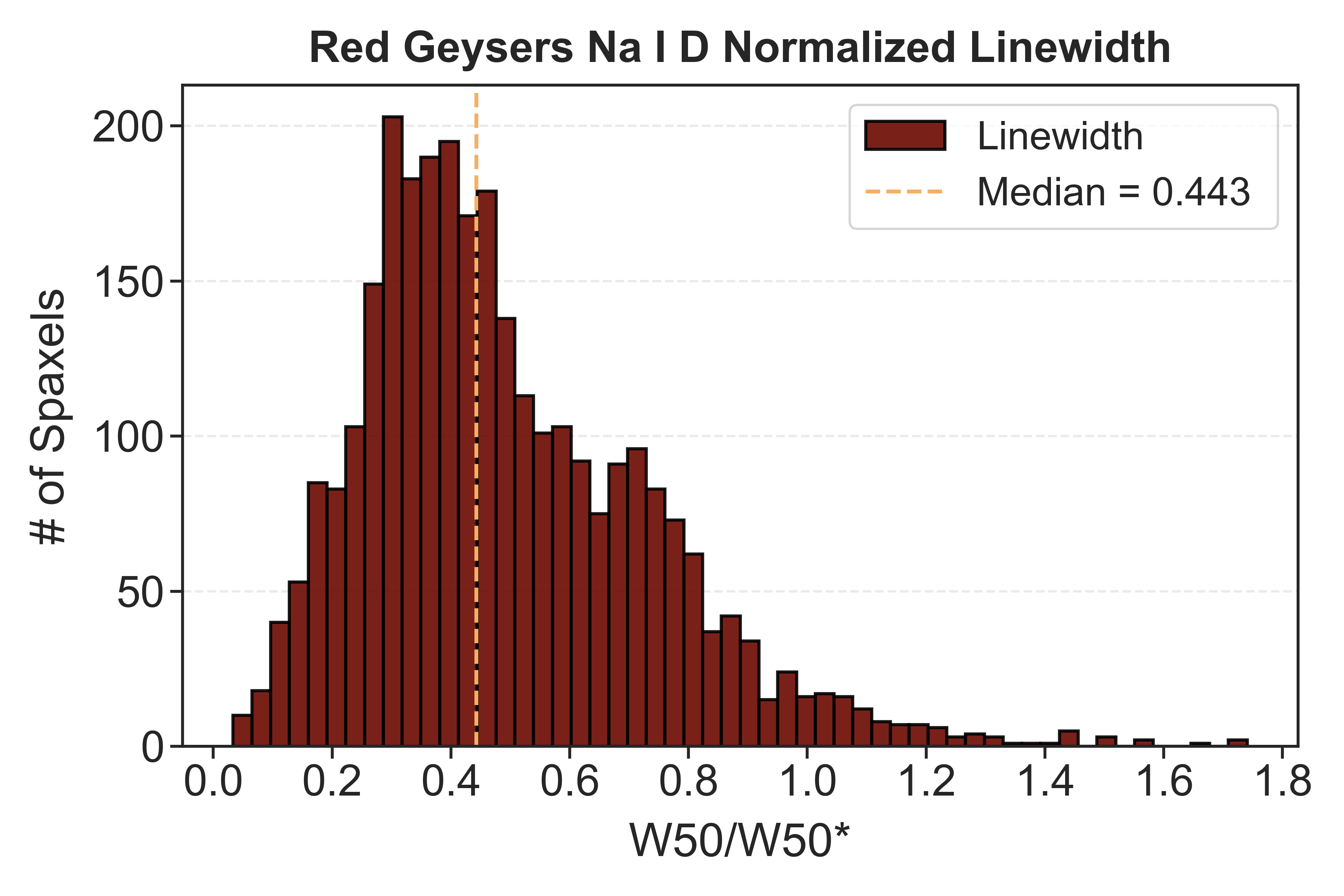} 
    \caption{Histogram of \NaID linewidth ratio $W_{50}/W_{50*}$ for spaxels with $S/N>5$ in red geyser galaxies.  Around $97\%$ of the normalized linewidth values are lower than 1, with a median of $0.443$, which indicates that these cool inflowing gases are dynamically colder than the stellar component in their host galaxy.}
    \label{fig:w50_hist}
\end{figure}

The histogram of $W50_{NaD}/W50_*$ across every spaxel of every red geyser galaxy in Figure~\ref{fig:w50_hist} demonstrates that the majority of spaxels lie at relatively low values, with a median $W50/W50*$ of $0.443$. We find that $\sim 97\%$ of all detected spaxels have a linewidth ratio $W50/W50* < 1$. The low velocity dispersion values of these cool, neutral gases indicate that relative to the stars in their host galaxies, these gas clouds are moving more coherently with less random motion.

\begin{figure}[h]
    \centering
    \includegraphics[width=0.48\textwidth]{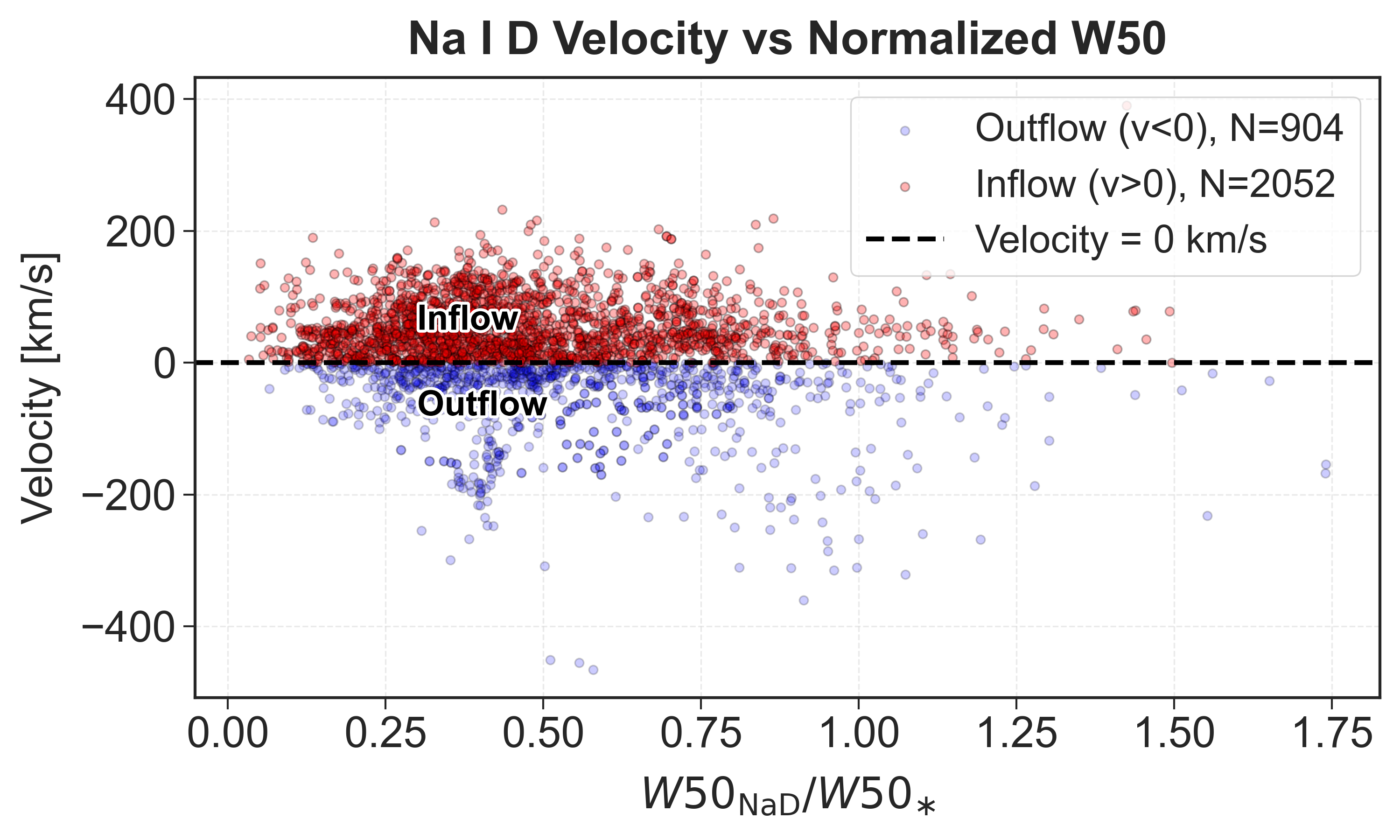} 
    \caption{\NaID velocity versus the normalized linewidth (with respect to the stellar linewidths) for all spaxels with a $S/N>5$ in red geyser galaxies. The red data points indicate inflowing gas relative to the host galaxy, while the less numerous blueshifted data points show the outflow. The horizontal dashed line indicates zero velocity. The figure shows that there are $\sim2$ times more inflowing \NaID detections compared to outflowing detections. Additionally, the linewidths of the inflowing and outflowing gas show no significant difference.} 
    \label{fig:velocity_linewidth}
\end{figure}

As a clear visualization of the relationship between these quantities, Figure \ref{fig:velocity_linewidth} shows the extracted \NaID velocities in red geysers plotted against their linewidths. Once again, the majority ($\sim 70\%$) of our detections are composed of inflowing gases, while the blueshifted spaxels make up $\sim 30\%$ of the sample. We also observe that across both inflowing and outflowing gases, absorption profiles are narrow, with median $W50/W50*$ values of $0.41$ for redshifted spaxels and $0.49$ for the blueshifted spaxels. Aside from their direction of motion, there is no significant difference between the kinematics of the inflowing and outflowing cool, neutral gases in red geysers.

\begin{figure}[h]
    \centering
    \includegraphics[width=0.48\textwidth]{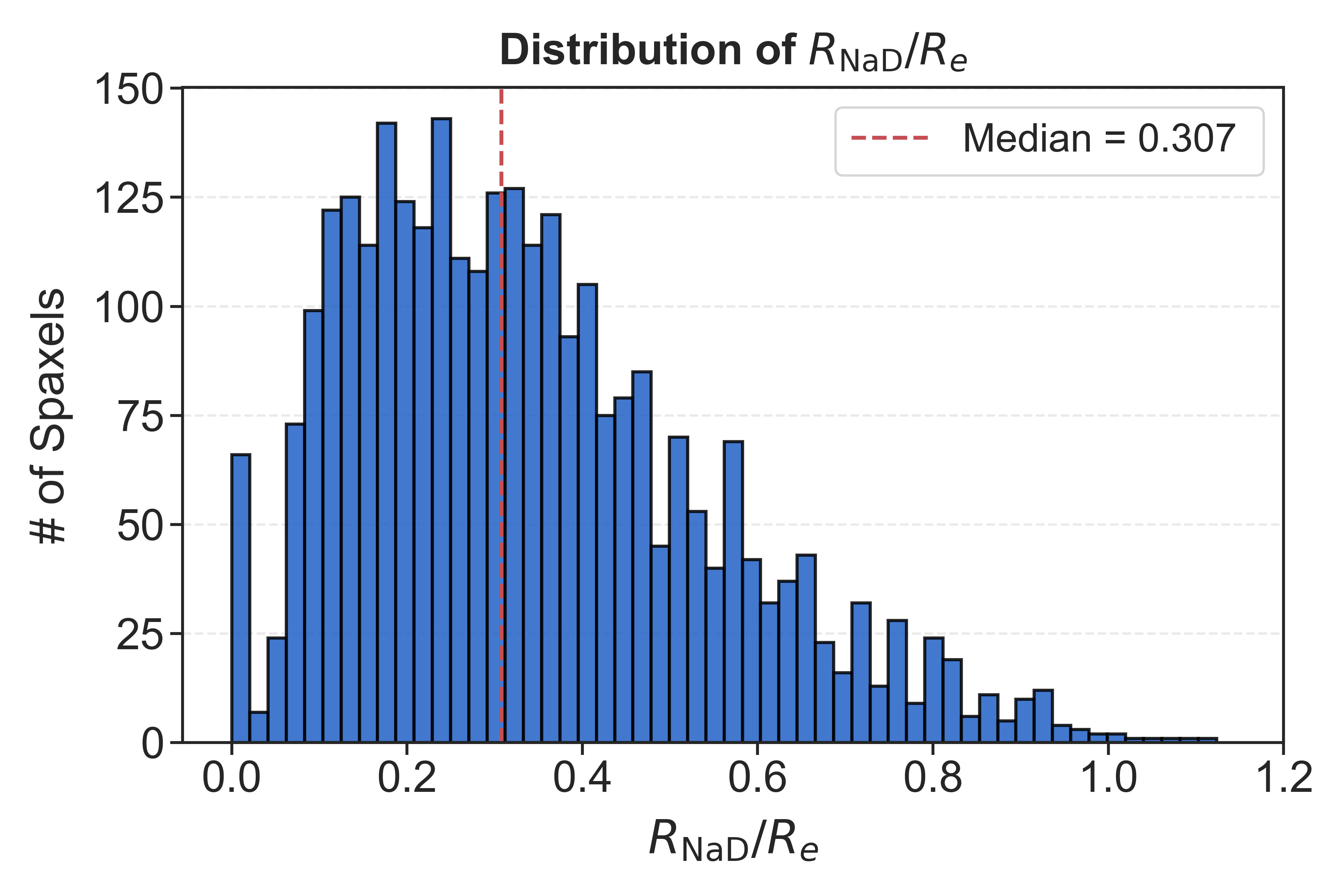} 
    \caption{Histogram of the projected distances of \NaID spaxels with a $S/N>5$, normalized by the host galaxy effective radius $R_e$. Nearly all detections are confined within 1 $R_e$, with a median of $R_{\text{NaD}}/R_e= 0.307$, indicating that the cool gas absorption is strongly concentrated in the central regions of red geyser galaxies.}
    \label{fig:R_NaD over R_e}
\end{figure}

We also calculate the distance from the center of each red geyser galaxy to the location of the \NaID detections, normalized by the effective radius of the host galaxy. As shown in Figure \ref{fig:R_NaD over R_e}, the cool, neutral gases are primarily concentrated in the central regions of the red geyser galaxy, with a median $R_{NaD}/R_e\approx0.3$. Only around $20\%$ of our detections come from regions beyond $0.5 R_e.$ In Subsection \ref{subsection:central concentration}, we explore potential detection biases and show that this central concentration is not due to detection sensitivity.

\begin{figure*}[t]
    \centering
    \includegraphics[width=1\textwidth]{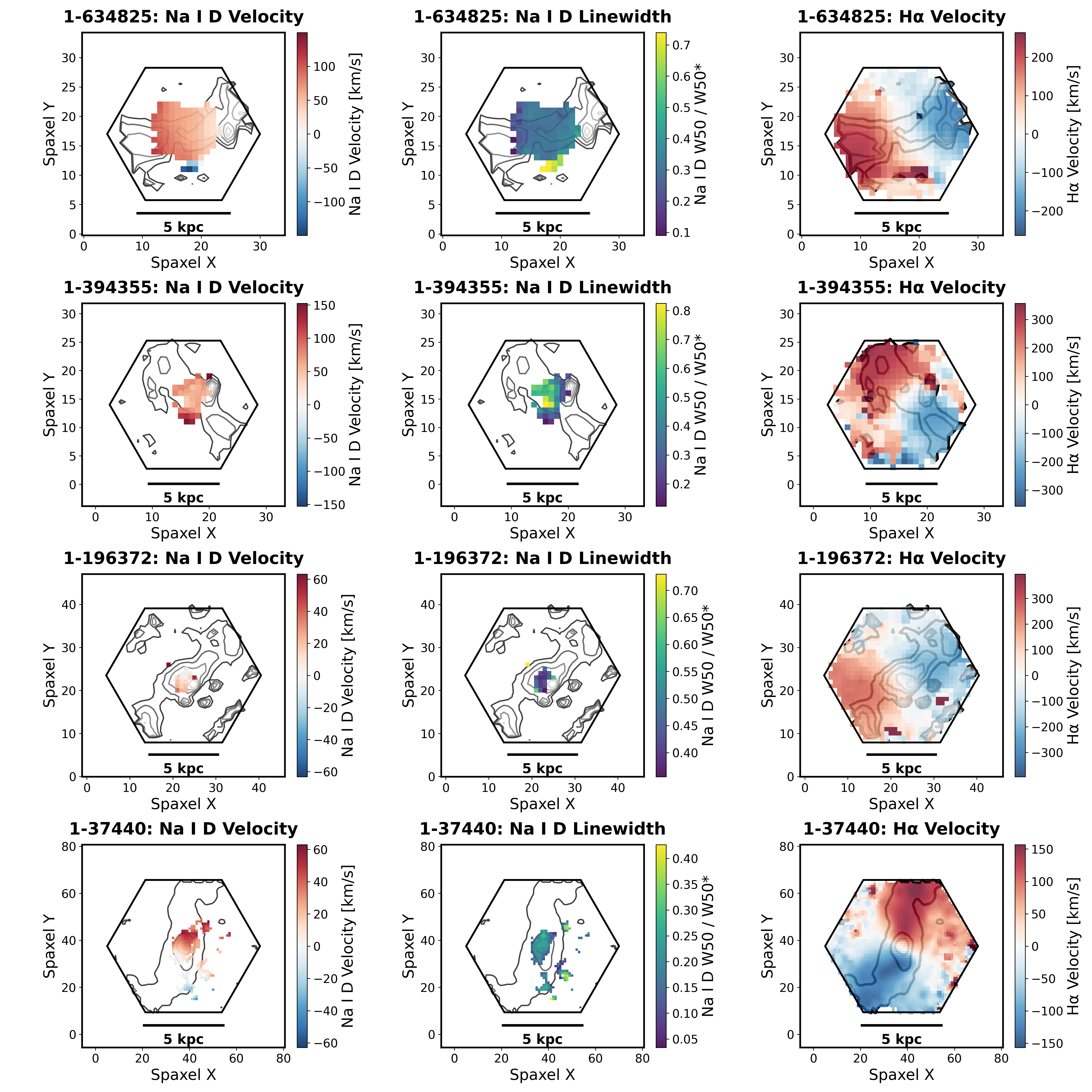} 
    \caption{\NaID kinematics maps for four red geyser galaxies (MaNGA IDs: 1-634825, 1-394355, 1-196372, 1-37440). In each map, the reliable spaxels ($S/N>5$) that are used in our analysis are shown with the contours of the top $\sim15\%$  $\text{H}\alpha$ EW in grey in the back to show the bi-symmetric features. The left column displays the velocities of these gases relative to the host galaxy, with red colors indicating inflowing motion, and the ($\sim2$ times less) blueshifted spaxels representing outflowing motion. The maps in the middle show the \NaID linewidth relative to the stellar component. In each case, \NaID absorption is predominantly concentrated in the central regions of the galaxies and does not seem to have any angular correlation with the outflowing bi-symmetric jet-like features. The kinematics show that these gases are mostly redshifted and are dynamically cold. The right column shows the $\text{H}\alpha$ velocities in red geyser galaxies, once again with red and blue representing inflows and outflows, respectively. There is no clear connection between the ionized gas kinematics and the cool, neutral gas motion traced by Na I D.}
    \label{fig:velocity_linewidth_maps}
\end{figure*}

To better understand the spatial distribution of the cool, neutral gas clouds, we create maps of our \NaID detections. A few of these maps are displayed in Figure \ref{fig:velocity_linewidth_maps}. Each row shows results for a different red geyser galaxy. The bi-symmetric features that define red geysers are depicted as black contours in each map, and on top, colored maps of the extracted \NaID velocities (left column), \NaID linewidths (middle column), and H$\alpha$ velocities (right column) are shown. 

Several key trends are apparent from these maps, which summarize our results:

(i) the \NaID detections are typically concentrated in the central parts of the red geyser galaxies---$80\%$  of the time within $0.5R_e$ of the host galaxy.

(ii) Around $70\%$ of the detected gas clouds are inflowing, lacking the symmetric inflow-outflow geometries observed in disks. These inflowing velocities are commonly around $47 \text{ km s}^{-1}.$ 

(iii) Compared to the stellar component, the \NaID absorption lines are relatively narrow and are typically found to have a $W50/W50*\sim0.4.$

\begin{figure}[h]
    \centering
    \includegraphics[width=0.45\textwidth]{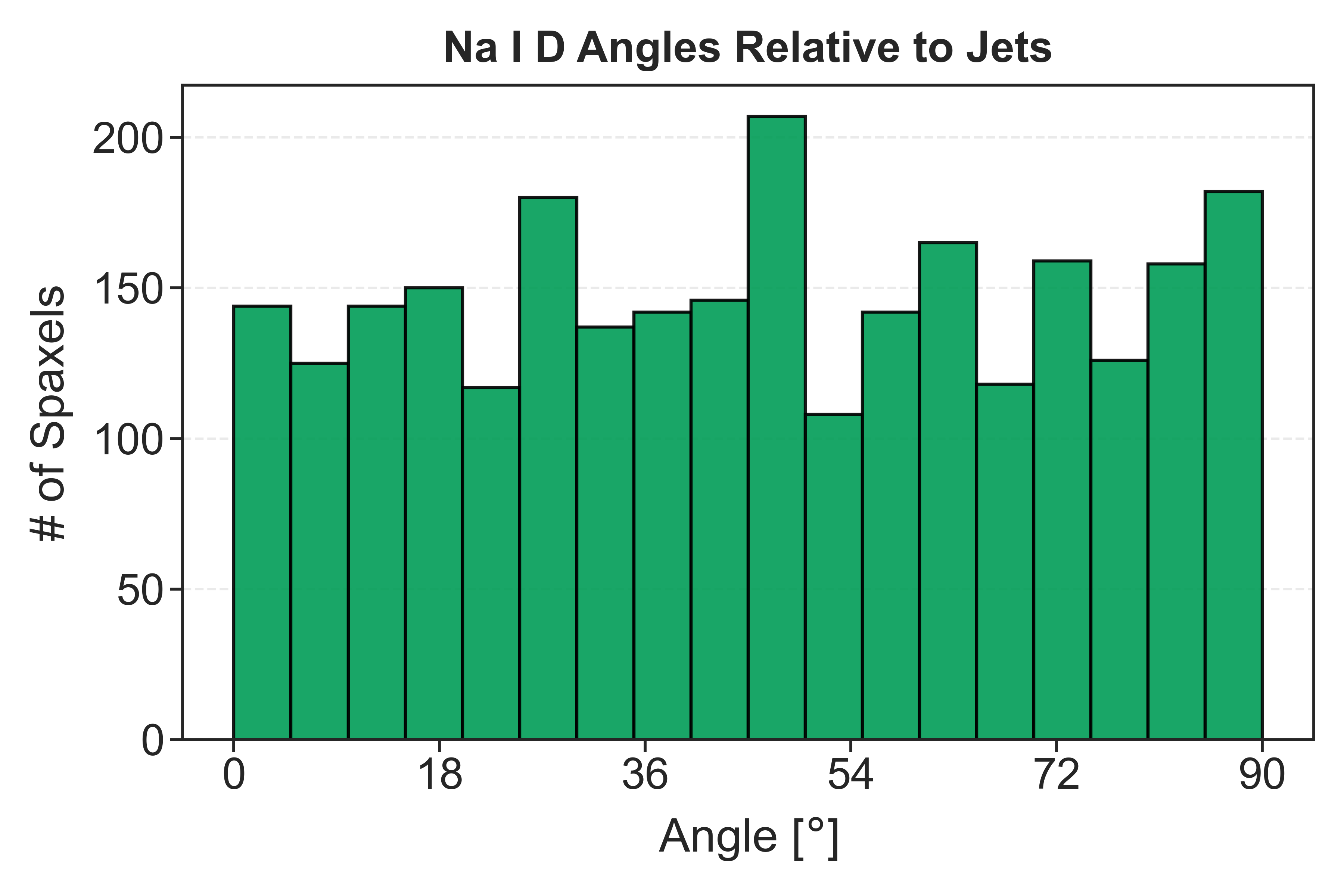} 
    \caption{Distribution of the angles between the spaxels with detected \NaID ($S/N>5$) and the H$\alpha$ EW bi-symmetric jet-like features across our sample of red geysers. There is no clear preferential direction for \NaID relative to the bi-symmetric features.}
    \label{fig:NaD & jets angles}
\end{figure}

(iv) From these maps, we also see that there seems to be no relationship between the \NaID detections and the bi-symmetric jets in $\text{H$\alpha$}$ EW maps. To better quantify this connection, first, we visually model the ``jets" with a straight line going from one side of the bi-symmetric $\text{H$\alpha$}$ EW feature at the edge of the MaNGA footprint, passing through the center, and reaching the opposite side. We then calculate the angle between the jet axis and a vector connecting the center of the line to each spaxel with detected ($S/N>5$) Na I D. Since the vector divides the jet axis into two angles on each side, we only take the smallest projected angle between the two directions, ranging from $0^\circ$ (aligned with the jet axis) to $90^\circ$ (perpendicular to the jet axis). A histogram of these angles within all red geyser galaxies is shown in Figure \ref{fig:NaD & jets angles}. There is no trend between the \NaID detections and the angle to the bi-symmetric features, meaning that the cool gas is directionally independent of the ionized outflows. However, as discussed earlier, the \NaID detections are primarily concentrated in the center, often remaining within the bi-symmetric outflows. Because of the limited spatial extent of these detections, our ability to test for angular correlations is restricted, and we could only be sensitive to extreme misalignments. 

Within these limits, we find no clear evidence for this cool, neutral gas to have preferential direction relative to the ionized outflows.


\subsection{\NaID Correlates with Radio Detection Status}
After applying quality cuts to remove spaxels with negligible \NaID absorption from the ISM, a large fraction of spaxels in each map were excluded (these can be seen clearly in Figure \ref{fig:velocity_linewidth_maps}). To quantify the cool, neutral gas available in the ISM of each red geyser galaxy, we use the retained spaxels to measure total projected area covered by the \NaID detections within each system. This area is a directly observable quantity that allows us to robustly analyze the relationship between cool gas, AGN activity, and environments of red geysers without requiring any additional assumptions. 

\begin{figure}[h]
    \centering
    \includegraphics[width=0.45\textwidth]{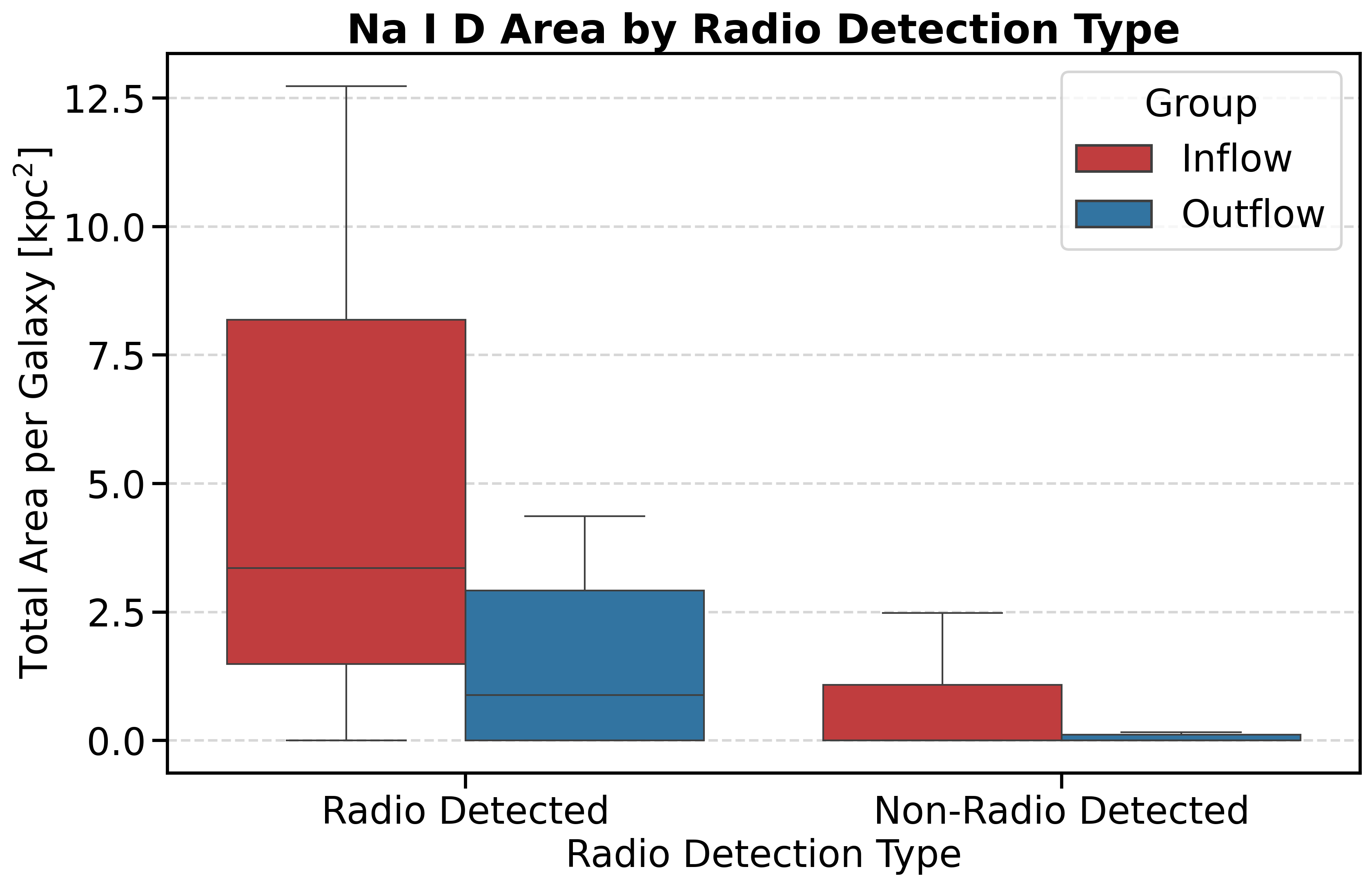} 
    \caption{Distribution of the projected area covered by \NaID detections with $S/N>5$ in each red geyser galaxy, grouped by inflow (red) and outflow (blue), for radio-detected and non-radio-detected systems. In total, there are 42 radio-detected and 98 non-radio-detected red geysers. The box plot shows that on average, the inflowing cool gases of radio-detected red geysers cover an area $\sim7$ times greater than their non-radio counterparts, with their median area of detections increasing from $0 \text{ kpc}^2$ to $\sim3.4 \text{ kpc}^2$. On the other hand, although the number of outflow detections remains low overall (going from a median detection area of $0 \text{ kpc}^2$ in non-radio to $\sim 0.9\text{ kpc}^2$ in radio-detected systems), their average area increases with AGN activity by a factor of $\sim 6$. The figure excludes outliers for better readability (based on the Interquartile Range rule).}
    \label{fig:boxplot radio}
\end{figure}

To test whether the spatial extent of \NaID absorption correlates with radio activity, we compare the area covered by the cool gas detections for radio-detected and non-radio-detected red geysers. We note that these subsamples have similar continuum S/N distributions, indicating that differences in detection sensitivity are likely not affecting the analysis. As shown in Figure \ref{fig:boxplot radio}, galaxies with radio detections tend to have more detected Na I D, particularly in the inflow group, with their area of detections jumping from a median of $0\text{ kpc}^2$ in non-radio red geysers to $\sim3.4 \text{ kpc}^2$ in radio-detected systems. We calculate that on average, the projected area of these inflowing gas reservoirs is around 7 times larger when a red geyser is radio-detected. In contrast, the area covered by outflow-classified spaxels remains low overall (with a median of $0\text{ kpc}^2$ in non-radio and $0.9\text{ kpc}^2$ in radio), but on average, their area increases by a factor of $\sim 6$. Indeed, a Mann-Whitney U test confirms that the sizes of the \NaID reservoirs differ on a statistically significant level among galaxies with different levels of AGN activity: for the inflows, we find that $p=3\times10^{-9}$, and for the outflows, we get $p=1.6\times10^{-6}$.

A good way to gain an intuition for these trends is to consider the fraction of galaxies in each category with inflowing \NaID detections that cover an area $> 1 \text{ kpc}^2$ in its host galaxy. In radio-detected systems, $\sim 76\%$ of the galaxies have inflowing \NaID with a projected area that is greater than $1 \text{ kpc}^2,$ compared to $\sim27\%$ of their non-radio counterparts. On the other hand, for blueshifted spaxels, roughly $48\%$ of radio-detected systems host outflowing \NaID with area $> 1\text{ kpc}^2$, while this fraction drops to $\sim 12\%$ among non-radio red geysers. 

Overall, increased AGN activity is associated with more extensive cool gas reservoirs.

\subsection{\NaID Correlates with Interaction Type}
\begin{figure}[h]
    \centering
    \includegraphics[width=0.45\textwidth]{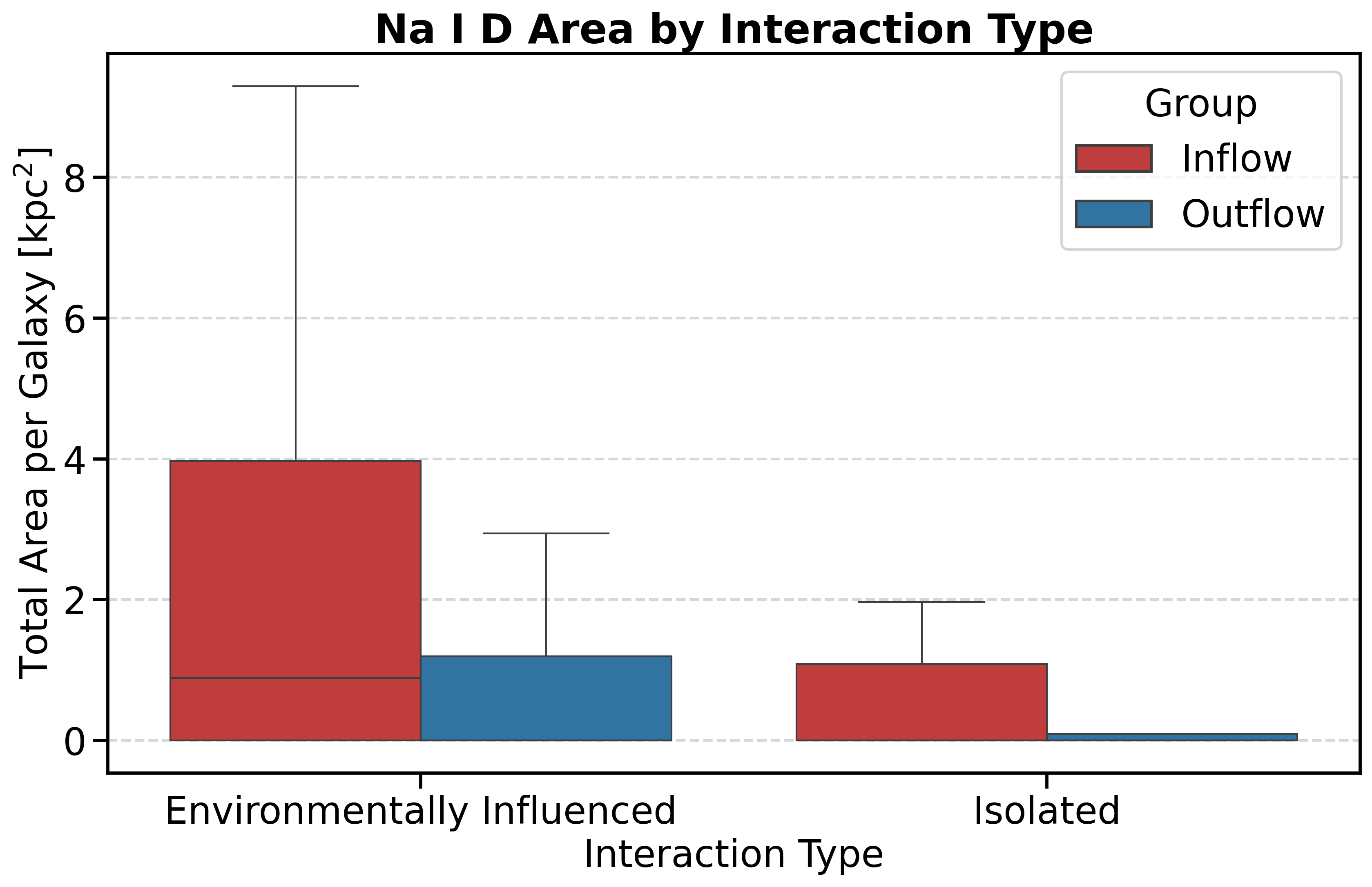} 
    \caption{Distribution of projected area covered by \NaID detections with $S/N>5$ in red geyser galaxies, grouped by inflowing (red) and outflowing (blue) gas, for two different galaxy interaction types. The box plot shows that on average, the environmentally influenced galaxies host inflowing gases over areas $\sim 2.7$ times greater than the isolated galaxies, with their median area of detections increasing with interaction from $0 \text{ kpc}^2$ in isolated to $0.9 \text{ kpc}^2$ in environment-affected systems. On the other hand, outflow detections remain low (consistently median 0) but their average area still increases by a factor of $\sim4.8$. Overall, through more interaction with nearby neighbors, red geysers show more widespread cool gases. The figure excludes outliers for better readability (based on the Interquartile Range rule).}
    \label{fig:boxplot interaction}
\end{figure}

A similar trend emerges when comparing the environmentally influenced and isolated red geysers. As shown in Figure \ref{fig:boxplot interaction}, the environmentally influenced galaxies tend to host more extensive inflowing cool gas reservoirs, on average covering an area $\sim 2.7$ times larger than that of their isolated counterparts, with their median area increasing from $0\text{ kpc}^2$ to $0.9\text{ kpc}^2$. Once again, our Mann-Whitney U test confirms that the distributions differ significantly across the interaction types: for inflows, $p=0.023$, and for outflows, $p=0.033$.

This suggests that interactions with nearby neighbors enhance the presence of inflowing \NaID absorption. On the other hand, the area of outflowing detections remain low (at a median of 0 $\text{ kpc}^2$) but still show a factor of 4.8 increase in their area when found in interactions or with nearby neighbors.

Another way to think about this result is through considering the fraction of galaxies in each group that host cool, neutral gas reservoirs with areas $> 1\text{ kpc}^2$. For red geysers that are categorized as environmentally influenced, $\sim50\%$ of their population show inflowing \NaID reservoirs with areas greater than $1\text{ kpc}^2$, compared to $25\%$ in the isolated systems. In contrast, outflowing, blueshifted spaxels are less common: our analysis shows that $27\%$ of the environmentally influenced red geysers host cool, neutral gas reservoirs with areas larger than $1\text{ kpc}^2$, but that fraction shrinks to $14\%$ for isolated galaxies. 

Overall, interaction with nearby neighbors, leads to more extensive cool gas reservoirs.

\section{\textbf{Discussion}}
\label{section:discussion}
We have studied the cool, neutral gases traced by the \NaID absorption lines in a sample of 140 red geyser galaxies, containing 42 radio-detected and 98 non-radio-detected galaxies. Around $\sim70\%$ of these galaxies are either clearly interacting, appear to be morphologically disturbed, or have a nearby neighbor, while $\sim30\%$ seem to be undisturbed and isolated.

First, our analysis shows that red geysers host systematically larger areas of the detected cool, neutral gas reservoirs compared to a control sample of
 galaxies. Specifically, the Na I D-detected areas in red geysers are $\sim 1.6$ times larger than the control sample (Figure \ref{fig:NaD Prevalence Histogram}. Furthermore, we show that the \NaID detection rate ($\geq5$ detected spaxels per galaxy) increases from $40\%$ in controls to $63\%$ in red geysers.

We also find that the cool, neutral gas clouds are predominantly (about $70\%$ of all detections) inflowing, with a median velocity of $\sim 47 \text{ km s}^{-1}$ (Figure \ref{fig:NaD velocities histogram}), and are moving coherently relative to the stars in the host galaxy with a median $W50/W50*\sim 0.4$ (Figure \ref{fig:w50_hist}). Additionally, we show that the \NaID is primarily concentrated in the central regions of red geyser galaxies, typically being found at an $R_{NaD}/R_{e}\sim0.3$ (Figure \ref{fig:R_NaD over R_e}). These cool, neutral gas clouds also show no preferred orientation relative to the ionized bi-symmetric outflow features that define red geyser galaxies. There is no trend between the \NaID detections and the angle to the bi-symmetric features, meaning that the cool gas is directionally independent of the ionized outflows. 

Historically, extensive studies of star-forming galaxies have reported blueshifted, outflowing cool gas absorption (\citealt{2009ApJ...692..187W}, \citealt{2010ApJ...719.1503R}, \citealt{2010ApJ...717..289S}, \citealt{2012ApJ...747L..26R}). However, it has been challenging to directly detect and study inflowing cool gas. 

Cosmological simulations, along with limited observational evidence, suggest that the inflowing cool gases may have relatively low metallicities and modest column densities and covering fractions (e.g., \citealt{2011MNRAS.418.1796F}, \citealt{2012ApJ...747L..26R}), which could suggest that the physical nature of the inflowing cool gas is different from the outflows that we commonly observe.

In this paper, we show that red geysers are dominated by inflowing cool gases. This property of red geysers could make them great laboratories for understanding the origin and the fate of inflowing cool gases in galaxies.

Interestingly, we show that red geysers that are radio-detected have more widespread regions of inflowing \NaID than their non-radio-detected counterparts, on average covering an area that is almost 7 times larger (Figure \ref{fig:boxplot radio}). 

In addition, we find that galaxy interactions further amplify this effect. Red geysers classified as ``environmentally influenced" show regions of appreciable inflowing \NaID absorption that are, on average, $\sim 2.7$ times larger than those in isolated, undisturbed systems (Figure \ref{fig:boxplot interaction}). Across both interaction and radio detection categories, the outflowing detections remain low.

Our results suggest that interactions between nearby galaxies can help with the import of cool, neutral gases. Although this is a good way of replenishing the cool gas reservoirs of these galaxies, interaction with nearby neighbors cannot be the only way, because only around 1 in 3 red geyser galaxies seem to be involved in an ongoing accretion process or a recent merger. The inflowing gas could also be a result of the cooling of the surrounding hot halo, which produces cool, neutral clumps of gas that rain down toward the galaxy center. There could also be a fountain scenario where the warmer outflowing gases cool and fall back down onto the host galaxies.

These gas clouds then fall to the centers of red geyser galaxies, feeding and sustaining the supermassive black hole activity in the center.

\subsection{Deviation from Free-Fall Velocities}

To see whether the observed inflowing gases behave like freely falling clouds in the host galaxy, we seek to calculate the free-fall velocities of the gas clouds. If we approximate the gravitational potential as one in which the circular velocity $v_{circ}$ is constant with radius, then the potential energy difference for a gas cloud moving from an initial radius $r_i$ to the current radius $r_f$ is
\begin{equation}
    \Delta U = m \int_{r_i}^{r_f} g(r) \, dr,
\end{equation}

where m is the cloud mass, and the gravitational acceleration is $g(r)=\frac{v_{circ}^2}{r}$. By evaluating the integral, we get $\Delta U=mv_{circ}^2\ln(\frac{r_i}{r_f})$. For this potential, $v_{circ}=\sqrt{2}\sigma,$ where $\sigma$ is the stellar dispersion. Thus, the free-fall velocity becomes 
\begin{equation}
    v_{ff} = 2\sigma \sqrt{\ln(\frac{r_i}{r_f})}.
\end{equation}

\begin{figure}[h]
    \centering

    \includegraphics[width=0.45\textwidth]{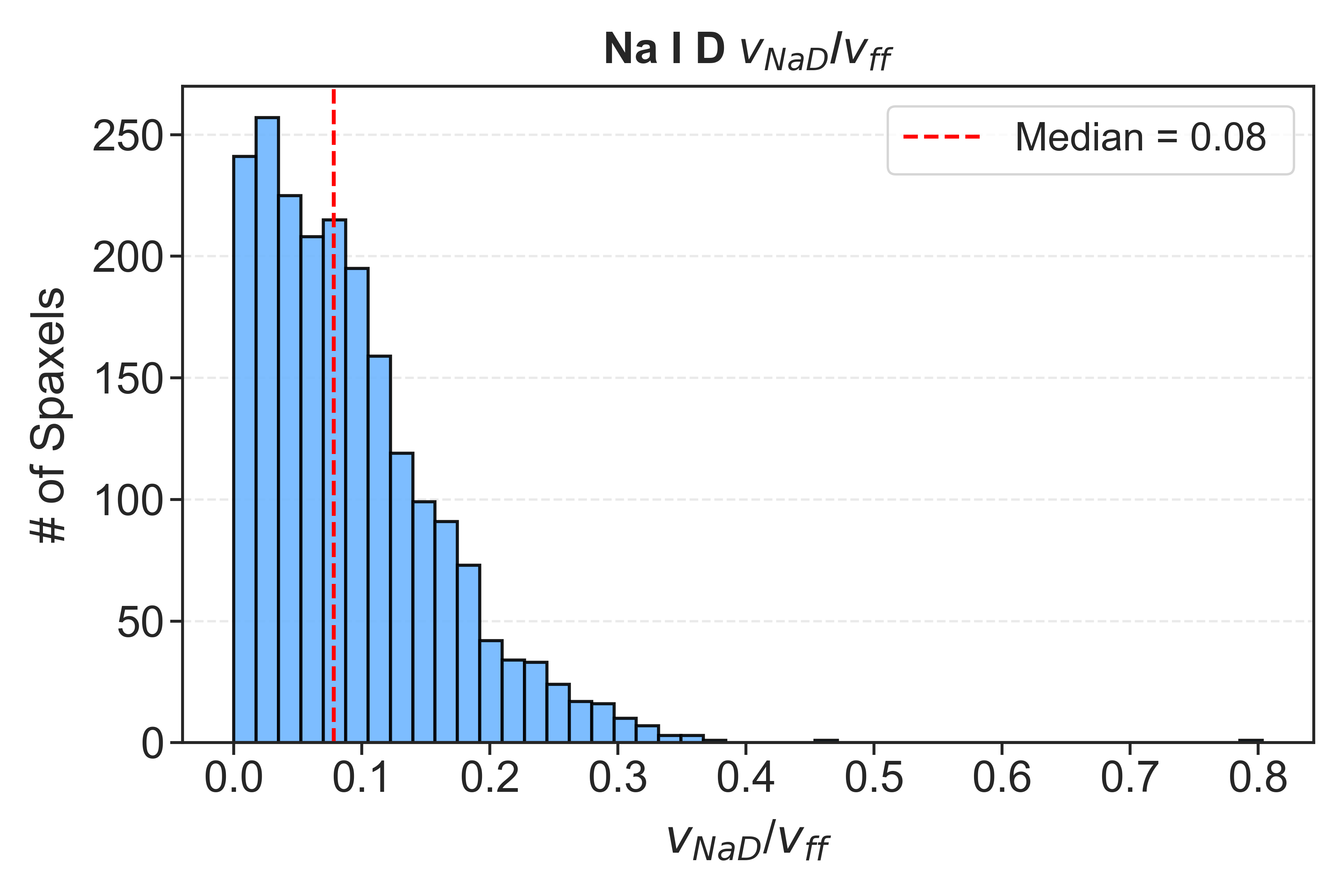} 
    \caption{Distribution of \NaID observed inflowing velocities relative to the calculated free-fall velocities of the gas clouds starting from $3R_e$ across spaxels with a $S/N>5$ in every red geyser galaxy. The cool, neutral gases traced by \NaID absorption are falling more slowly than we would expect in free-fall, with a median $v_{NaD}/v_{ff}\sim 0.1.$}
    \label{fig:v_NaD_over_vff}
\end{figure}

We compare the measured \NaID velocities to the expected free-fall velocity starting from $\text{3R}_e.$ In the case in which the cool gas is freely falling toward the center under the gravitational potential of its host galaxy, we would expect $\frac{v_{NaD}}{v_{ff}}\approx1$. However, in Figure \ref{fig:v_NaD_over_vff}, where we have normalized the observed \NaID velocities with the calculated free-fall velocities across the spaxels with a $S/N>5$ in all red geyser galaxies, we show that the ratios are well below unity, with a median of $\sim 0.1$. 

This discrepancy could partially be explained by line-of-sight effects, which on average reduce the true 3-D velocity by a factor of $\sqrt{3}$. An additional factor in play could be the presence of drag forces from the hot halo gas. The idea is that as the cool gases fall through the hot gas, they could experience ram pressure and slow down. In such a case, the cloud would asymptotically approach a terminal velocity defined by the condition that the gravitational and drag forces balance:
\begin{equation}
    \rho \ v_{term}^2 \ A_{cloud} = M_{cloud} \frac{v_{circ}}{r}.
\end{equation}
Here, $\rho$ is the density of the surrounding hot gas, $A_{cloud}$ is the cross-sectional area of the cloud, and $M_{cloud}$ is its mass. We can write 
\begin{equation}
    M_{cloud} = A_{cloud}  N_H  <\mu>  m_p,
\end{equation}
with $N_H$ as the hydrogen column density ($\text{cm}^{-2}$), $<\mu>\approx1.4$ as the mean atomic weight per hydrogen nucleus (accounting for helium), $m_p$ as the proton mass, and $r$ as the galactocentric distance of the cloud from the center. Substituting in our equations and rearranging, we get the following:
\begin{equation}
    v_{term}=\sigma\sqrt{\frac{2N_H <\mu> m_p}{\rho r}}.
\end{equation}

To estimate $v_{term}$ for \NaID absorbers, we therefore require both a reasonable estimate of the hot gas density $\rho$ at the radial distance of the detected gases and the hydrogen column density $N_H$.

To estimate the density of the hot gas, we adopt a standard $\beta$-model:
\begin{equation}
    n_e(r) = n_{e,0} [1+(\frac{r}{a_x})^2]^\frac{-3\beta}{2},
\end{equation}
where $n_{e,0}$ is the central electron density, $a_x$ is the core radius, and $\beta$ is the slope parameter. Based on the data in \citet{1985ApJ...293..102F}, we assume that $n_{e,0}=10^{-2}\text{cm}^{-3},$ $a=3\text{ kpc},$ and $\beta=0.5.$ From this electron density profile, we extract the corresponding mass density:
\begin{equation}
    \rho(r) = \mu\ m_p\ n_e(r),
\end{equation}
where $\mu$ is the mean molecular weight per free electron and is set to be 1.18.

Our calculations show that the terminal velocities are an order of magnitude higher than the \NaID velocities observed, suggesting that drag from hot gases is likely not playing a major role in slowing down these cool gas clouds.

These modest inflow velocities and the absence of \NaID absorption at large radii, in addition to the fact that $\sim 2/3$rd of the red geysers show no signs of interaction, suggest that the observed neutral gas does not always originate directly from large-scale accretion or interactions but could also originate from condensation and cooling of ionized material in the inner few kilo-parsecs. This scenario could more naturally explain both the restricted spatial extent of the absorption and the low velocities. 

Using $g=\frac{v_{circ}^2}{r}$, we estimate the time to reach the observed velocities under gravitational acceleration: 
\begin{equation}
    t_{accel} = \frac{v_{obs}}{g}.
\end{equation}

We calculate that if these clouds start falling inward under gravity, they would reach their observed velocities after $\sim1$ Myr. This very short timescale implies that the clouds are young and have not had sufficient time to accelerate to full free-fall velocities, consistent with our finding that $v_{obs}<v_{ff}$. This suggests that the Na I D-traced gas represents a temporary phase of cool gas, likely formed recently through condensation from the ionized phase.

\subsection{Central Concentration of \NaID}
\label{subsection:central concentration}
As described in \S\,\ref{sec:results}, the detected \NaID absorbers are centrally concentrated with a median $\frac{R_{NaD}}{R_{e}}\sim0.3.$

To determine whether this concentration of the \NaID detections in the center is a genuine physical feature or a detection bias caused by higher signal-to-noise and continuum flux near the galaxy centers, we examine how the threshold for a reliable $5\sigma$ equivalent width (EW) detection---estimated using the Monte Carlo method described in \S\,\ref{subsection: gaussian fitting}---depends on the continuum flux. Then, we measure how the continuum flux changes with radial distance from the galaxy center. Combining these relationships, we then derive the dependence of the $EW_{5\sigma}$ on the distance of the cool gas detection from the center. 

\begin{figure}[h]
    \centering
    \includegraphics[width=0.45\textwidth]{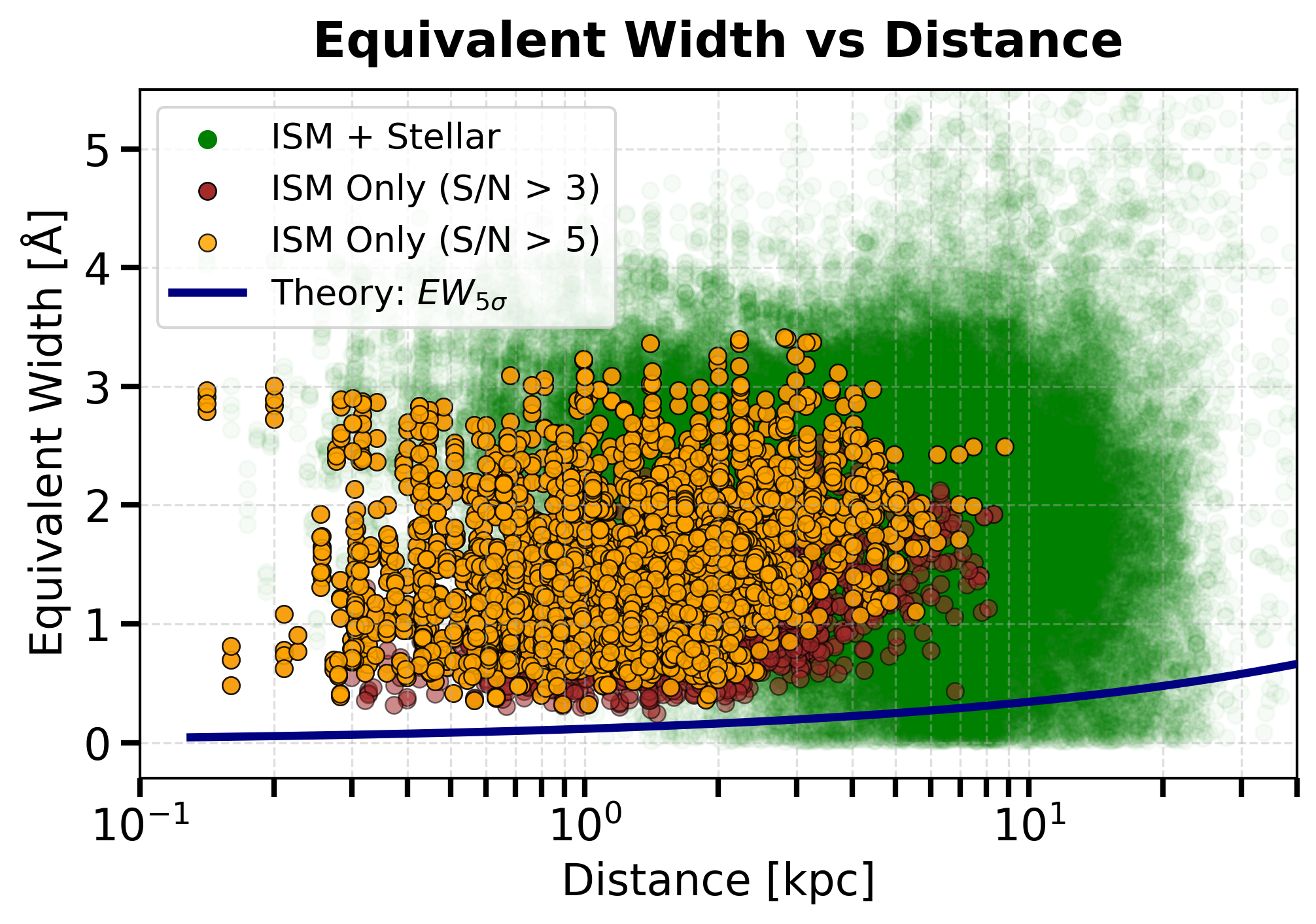} 
    \caption{Equivalent Width (EW) of \NaID absorption as a function of distance from the center of the galaxy. The green data points show all detected \NaID features (including both interstellar medium and stellar contributions), the yellow points indicate detections after removing the stellar component using the DAP and cutting out spaxels with a $S/N<5$, and the red points correspond to the ISM detections after removing spaxels with a $S/N<3$. The blue curve represents the modeled $EW_{5\sigma}$ detection threshold derived from the dependence of the continuum flux on radial distance from center. The comparison demonstrates that the observed \NaID equivalent widths lie well above the $5\sigma$ detection limit across all radii and confirms that the measured absorption features, especially their concentration near the center, are not limited by detection sensitivity. Additionally, while the S/N cuts increase the number of data points, they do not significantly extend our detections to the outskirts of their host galaxies.}
    \label{fig:ew_radial_constraint}
\end{figure}

In Figure \ref{fig:ew_radial_constraint}, we plot the \NaID EW detections before and after removing the stellar component and cutting out spaxels with $S/N<5$ and $S/N<3$. We also plot our calculation for how a $5\sigma$ detection is expected to change as a function of radial distance from the center and find that our \NaID detections lie above the theory $EW_{5\sigma}$ and are not limited by the detection sensitivity. However, the decreasing number of detections with low EW values in the outskirts may still reflect decreasing sensitivity or lower S/N at large radii. 

To test this possibility, we repeat our analysis with a less restrictive $S/N$ cut of 3. Our results show that softening the $S/N$ threshold increases the number of available spaxels by a factor of $1.5$, but it does not significantly extend our detections beyond current radii. This suggests that although we might be missing cool gas with weak absorption in the outskirts, the decline in the number of \NaID detections could reflect a genuine decrease in the abundance of cool, neutral gases in the outer parts of red geysers.

\subsection{Calculating \NaID Properties}

In order to estimate the total gas mass and inflow rate in red geyser galaxies, we require the hydrogen column density, which we can get from the sodium column density $N(Na I)$ below:
\begin{equation}
\label{equation:N_NaI}
   N(NaI)= \frac{\tau_0\ b}{1.497 \times10^{-15}\lambda_0 f}, 
\end{equation}

where $\tau_0$ and $\lambda_0$ are the central optical depth and central wavelength of each line component, respectively, $b$ is the Doppler linewidth ($b=\sqrt{2}\sigma=\frac{W50}{2\sqrt{\ln 2}}$), and $f$ is the oscillator strength. Throughout our calculations, we choose $\lambda_0 = 5897.55\text{\AA}$ and $f=0.318$ \citep{1991ApJS...77..119M}.

\begin{figure}[h]
    \centering
    \includegraphics[width=0.45\textwidth]{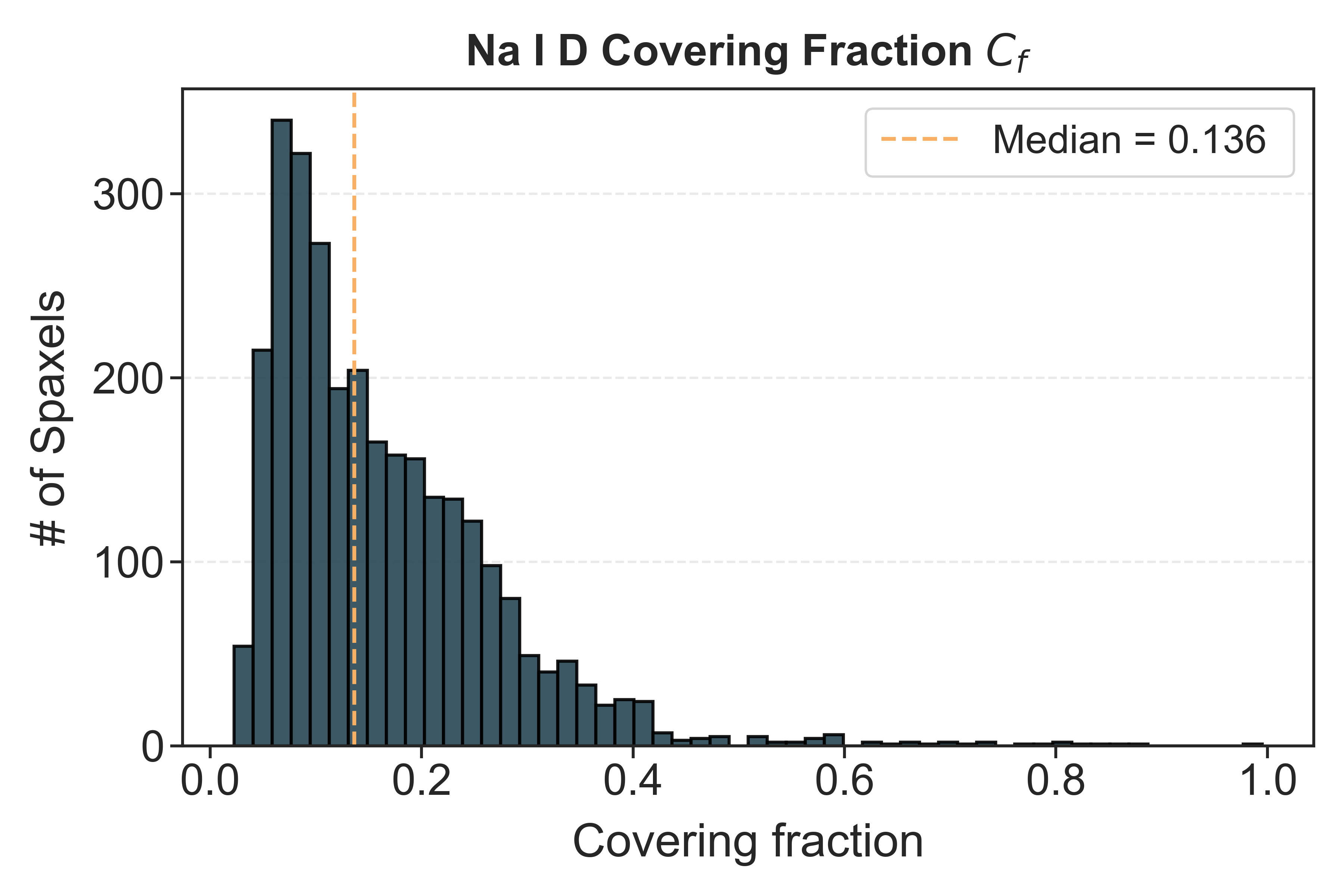} 
    \caption{Covering fractions of \NaID across spaxels with a $S/N>5$ in every red geyser galaxy. The median of the distribution, shown using an orange dashed line, is at $0.14$, implying the presence and dominance of small and patchy absorbing gas clouds in red geyser galaxies.}
    
    \label{fig:covering_fractions}
\end{figure}

To get the optical depth, we first calculate the covering fraction following \citet{1997ASPC..128...19H}, \citet{1997AJ....113..136B}, and \citet{2000ApJS..129..493H}, 
\begin{equation}
\label{equation:covering_fraction}
    C_f=(I_{R}^2 - 2I_{R} + 1)/(I_{B} - 2I_{R} + 1),
\end{equation}
where $I_{R}$ is the normalized intensity at the center of the D1 line, and $I_{B}$ is the normalized intensity at the center of the D2 line. The distribution of the covering fractions of the \NaID across our sample of red geysers is shown in Figure \ref{fig:covering_fractions}. From the distribution, we calculate that these cool neutral gases in red geysers have a median covering fraction of $\sim0.14$, indicating that the absorbing gas is highly clumpy and covers a small fraction of each MaNGA spaxel. In fact, we show that $\sim 98\%$ of the detections have a covering fraction smaller than $0.5.$

Continuing our calculation, once again following \citet{1997ASPC..128...19H},
\begin{equation}
    \tau=\ln [C_f / (I_{5896} + C_f-1)].
\end{equation}

\begin{figure}[h]
    \centering
    \includegraphics[width=0.45\textwidth]{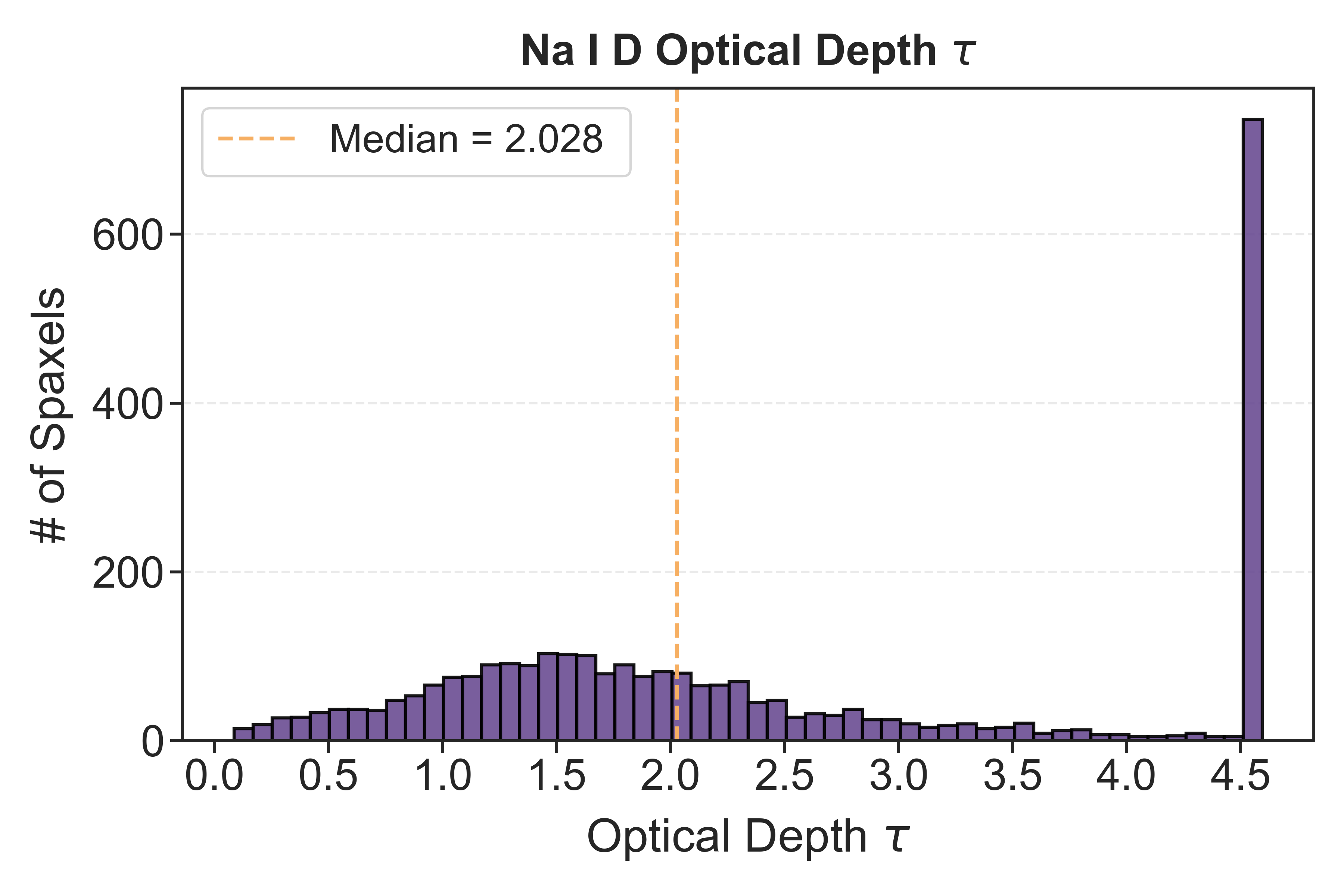} 
    \caption{Distribution of \NaID optical depths $\tau$ measured in all spaxels with a $S/N>5$ across the full red geyser sample. The histogram shows two distinct features: (i) a low-$\tau$ peak with a median of $\tau\sim1$, which makes up the majority ($75\%$) of the detections and represents weakly absorbing gases; (ii) a sharp high-$\tau$ spike at $\tau>4$ that makes up $25\%$ of the sample. The vertical dashed orange line shows the median optical depth at $\tau\approx2$, which indicates that a large fraction of the cool neutral gases in red geysers is in the optically thick regime. Together, these features suggest that the \NaID absorption gets contributions from gas clouds with a range of different optical depths, especially those that are more optically thick.}
    
    \label{fig:optical_depth}
\end{figure}

A summary of our optical depth measurements across the spaxels ($S/N>5$) of every red geyser is shown in Figure \ref{fig:optical_depth} with a median of $\tau\approx2$. We find two major populations of optical depth detections: (i) detections with a median $\tau\sim1.5$, suggesting the presence of a small population of more diffuse neutral gas throughout these systems, and (ii) detections with $\tau>4$, which make up $\sim 25\%$ of the entire sample, suggesting a significant population of very opaque gas clouds within red geysers. The coexistence of widespread low-opacity absorption and extremely optically thick spaxels strongly suggests that these quiescent galaxies host a multiphase reservoir of cool neutral gas rather than a single, uniform gas supply.

\begin{figure}[h]
    \centering
    \includegraphics[width=0.45\textwidth]{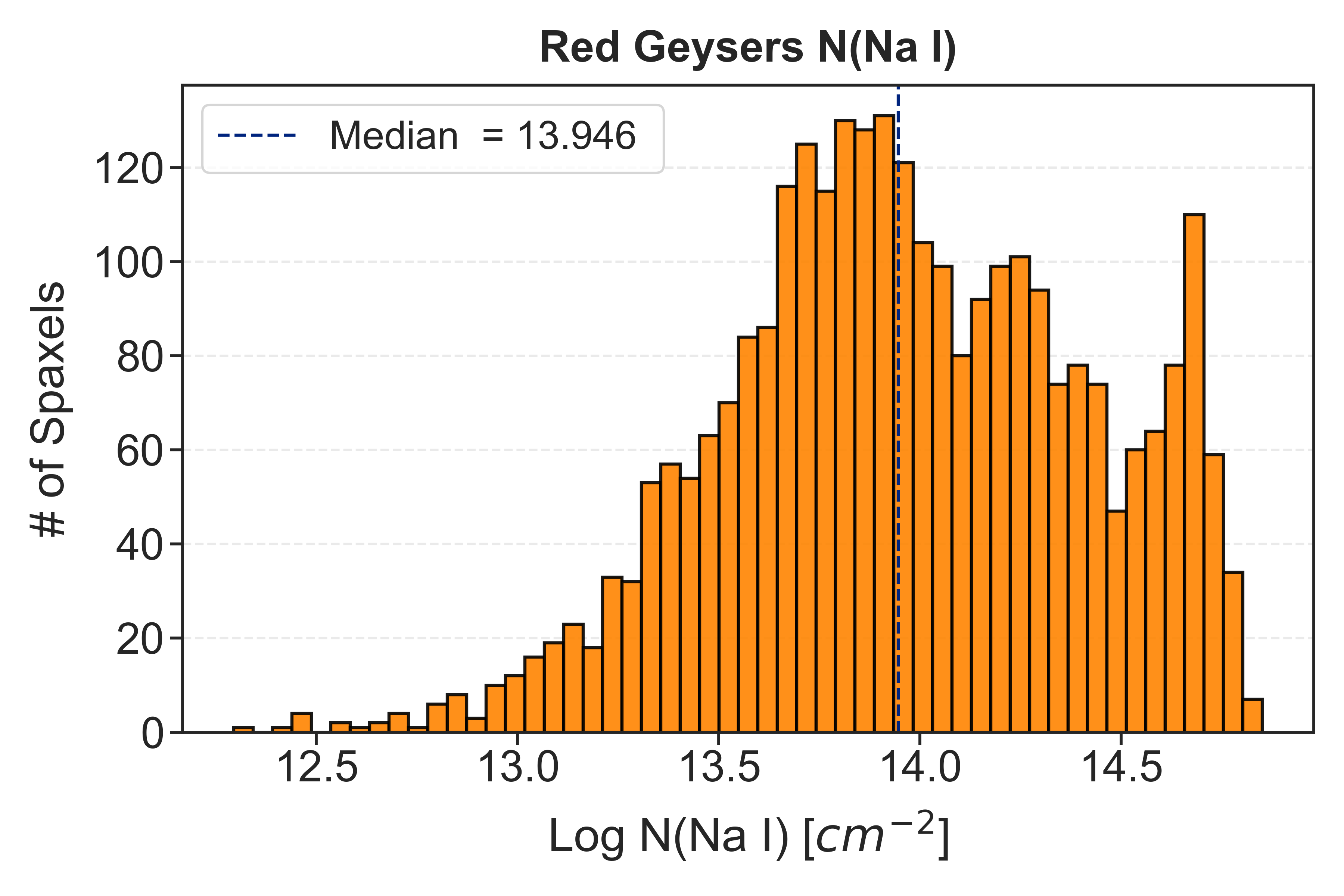} 
    \caption{Distribution of the Na I column density in spaxels with a $S/N>5$ across the full red geyser sample. The distribution peaks around the median value of $\log N_{NaI}\sim 13.95$, shown using the blue dashed line. Its left tail toward low column densities indicates a small population of cool, neutral gases at low densities (20th percentile at $\log N_{NaI}\sim 13.6$), while the extended high-column density on the right demonstrates the presence of a large number of dense areas (80th percentile at $\log N_{NaI}\sim 14.4$). The neutral sodium column densities span almost 3 orders of magnitude, demonstrating the wide range of cool, neutral gas environments present in red geysers.}
    
    \label{fig:N_NaI_histogram}
\end{figure}

By estimating the central optical depth $\tau_0$ and the Doppler parameter $b$, we can finally use Equation \ref{equation:N_NaI} to determine the sodium column density $N(NaI)$. In Figure \ref{fig:N_NaI_histogram}, we have calculated the neutral sodium column density for all spaxels with $S/N>5$ across the full sample. The median is measured to be at $\log N_{NaI}\sim 13.95$, while the 20th percentile of the distribution is at $\log N_{NaI}\sim 13.6$ and its 80th percentile at $\log N_{NaI}\sim 14.4$. The large range of neutral sodium column densities (nearly 3 orders of magnitude) suggests a large absorption contribution from both dense and diffuse structures.

To convert the sodium column densities to hydrogen column densities $N_{\rm H}$, we use empirical data from \citet{1995ARA&A..33...19H}, \citet{1977ApJ...216..291S}, and \citet{1985ApJ...298..268S}, where the $N_{\rm NaI}$ and $N_{\rm H}$ of many different Milky Way sightlines are given, to extract a relationship between the two column densities, which we find to be the following: 
\begin{equation}
\label{equation:N_H}
    \log N_\mathrm{H} \approx 0.29 \, \log N_\mathrm{Na\,I} + 17.21.
\end{equation}

\begin{figure}[h]
    \centering
    \includegraphics[width=0.45\textwidth]{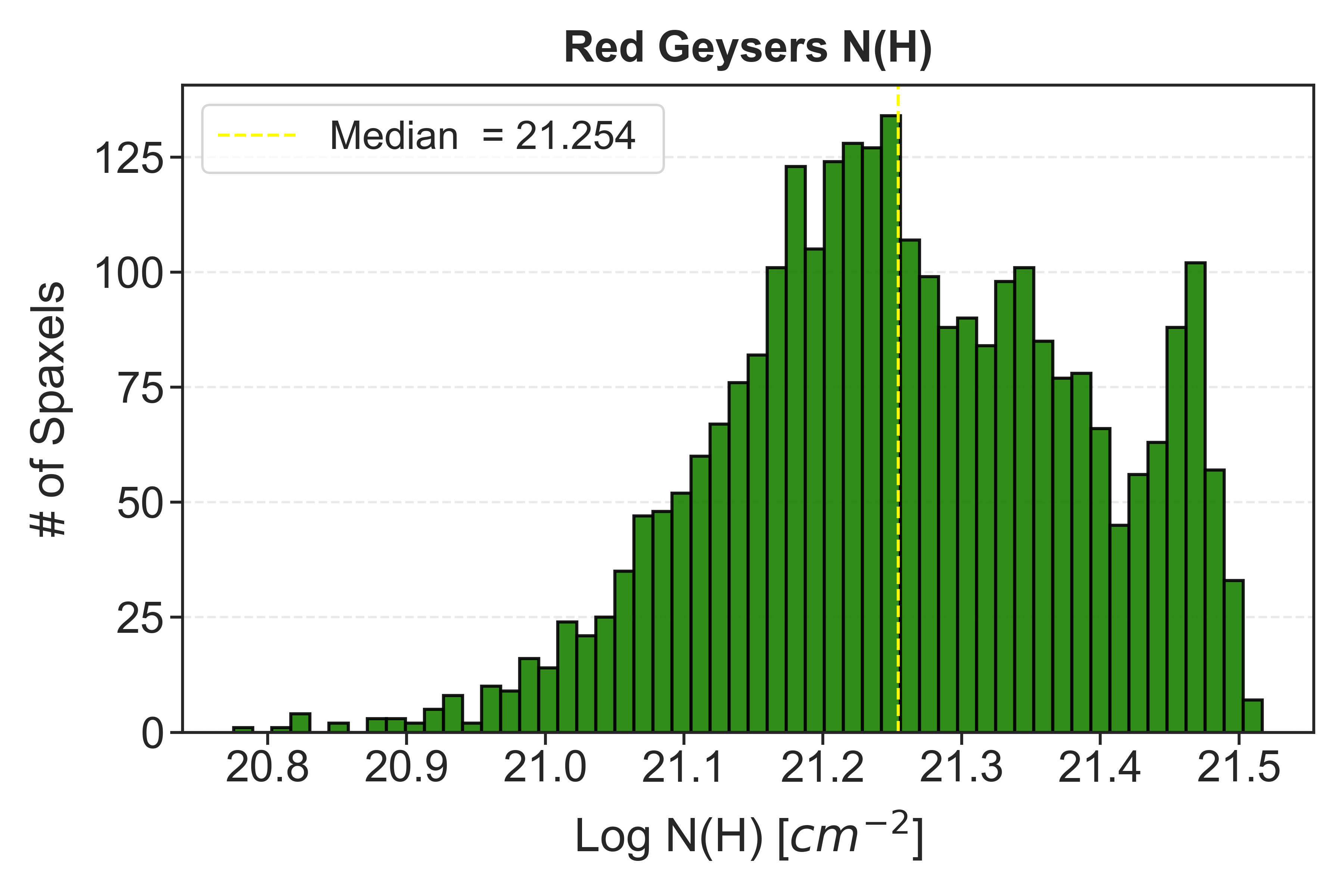} 
    \caption{Distribution of the hydrogen column density in the spaxels with a $S/N>5$ across the full red geyser sample. The median is shown with the yellow dashed line at $\log N_{H}\approx 21.25$. The histogram extends less than one order of magnitude, suggesting less diversity in hydrogen gas environments compared to the neutral sodium.}
    
    \label{fig:N_H_histogram}
\end{figure}

Using this relation, we calculate the hydrogen column densities in all spaxels with an \NaID $S/N>5$ in every red geyser galaxy, as shown in Figure \ref{fig:N_H_histogram}. The median is found to be $\log N_{H} =21.254$, and the entire sample spans less than one order of magnitude, compared to $\sim 3$ orders of magnitude found in neutral sodium column densities, implying that while the cool neutral sodium exhibits a large range of column densities from spaxel to spaxel, the total hydrogen reservoir is relatively uniform.

\begin{figure}[h]
    \centering
    \includegraphics[width=0.45\textwidth]{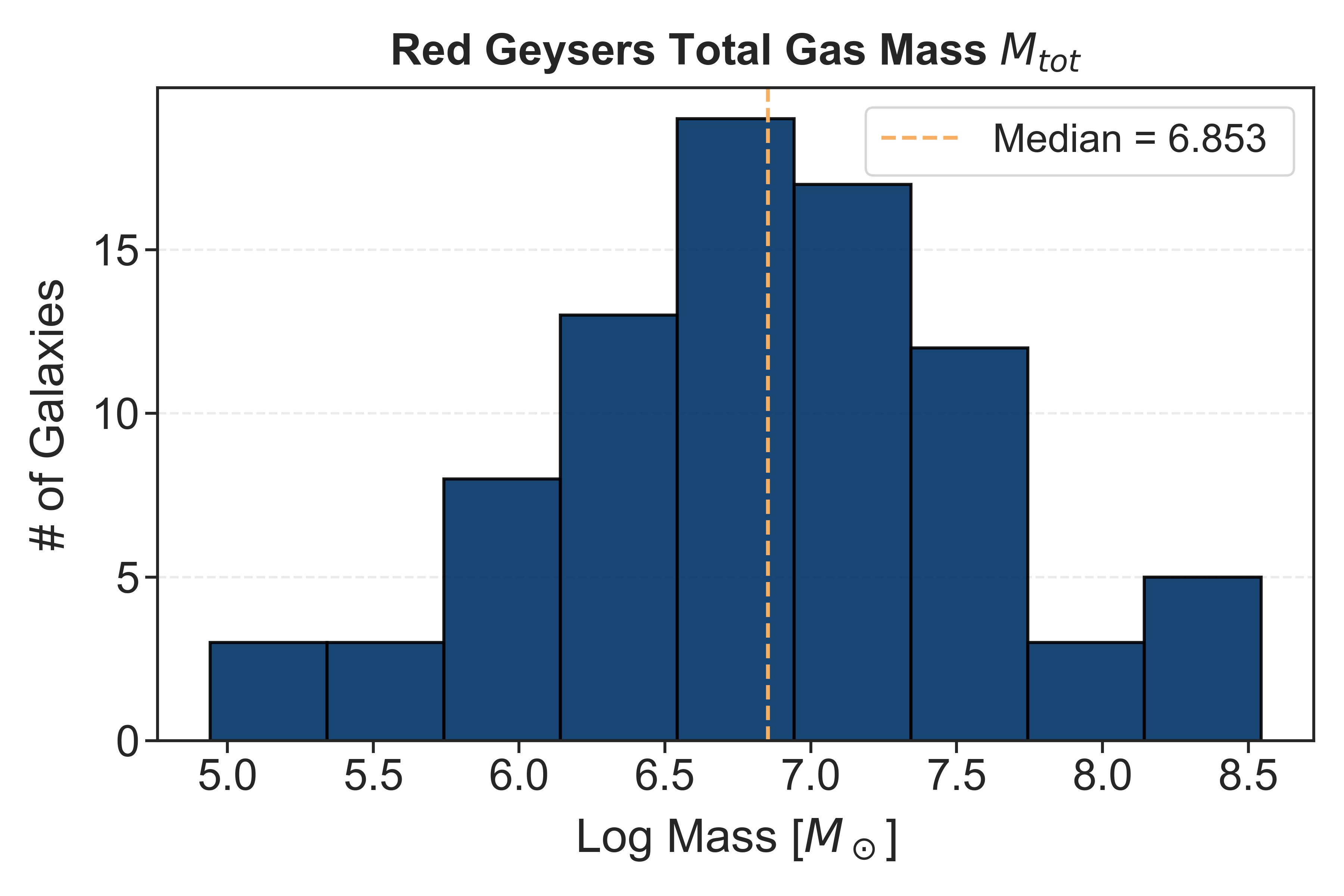} 
    \caption{Distribution of the total (inflowing+outflowing) gas mass $M_{\rm tot}$ in each red geyser galaxy, including only galaxies with detected \NaID spaxels at a $S/N>5$ ($60\%$ of the full sample). The distribution extends from $\sim 10^5 - 10^8M_\odot$, and the median, shown in yellow, is found to be at $\log M_{tot}\approx6.85$. These results suggest that while $40\%$ of red geysers are gas poor, the rest of the sample hosts modest gas reservoirs with a diverse range of masses. 
    }
    \label{fig:total_gas_mass_distribution}
\end{figure}

Using the hydrogen column density $N_H$, we can now extract the inflow rate of the gases and their total mass. We calculate the total mass as 
\begin{equation}
    M_{tot} = m_H \cdot \mu \cdot C_f \cdot N(H) \cdot A_{NaD},
\end{equation}

where $m_H$ is the mass of the Hydrogen atom, $\mu = 1.4$ accounts for the contribution of helium to the total gas mass, $C_f$ is the covering fraction extracted from Equation \ref{equation:covering_fraction}, N(H) is the hydrogen column density calculated in Equation \ref{equation:N_H}, and $A_{NaD}$ is the physical projected area covered by the detected \NaID absorption. A histogram of the calculated gas masses in red geyser galaxies with at least one reliable ($S/N>5$) spaxel, which is $\sim 60\%$ of the sample, is shown in Figure \ref{fig:total_gas_mass_distribution}. The median gas mass is found to be $\sim10^7 M_\odot$, and the entire distribution spans more than three orders of magnitude. In total, when including the galaxies with no \NaID detections, the median comes down to $\sim 10^6  M_\odot$. Our result suggests that although about $40\%$ of the red geysers are gas poor, the rest contain modest amounts of gas with a wide range of masses. 

The mass inflow rate can be related to the total gas mass and the accretion timescale by $\dot{M} \sim M_{\rm tot}/ \ t_{\rm acc}$, where the accretion time $t_{\rm acc}\sim\frac{r}{v},$ in which $r$ represents the projected radial extent of the \NaID traced gas from the center, and $v$ is the measured \NaID velocity relative to the host galaxy.

The accretion timescale provides a rough estimate for the time it takes these clouds to reach the center at their observed velocities, thus providing a sense of their lifetimes. We calculate the median accretion time to be $\sim 2\times 10^7\text{ yr}$, with the 16th and 84th percentiles corresponding to $\sim10^7 \text{ yr}$ and $\sim 8\times10^7 \text{ yr}$. These relatively short lifetimes once again show that these clouds are short-lived and likely destroyed within short timescales.

Given the accretion timescales and the total gas mass, we can now calculate the mass inflow rate:

\begin{equation}
    \dot{M} = M_{tot}\frac{v}{r}.
\end{equation}

\begin{figure}[h]
    \centering    \includegraphics[width=0.45\textwidth]{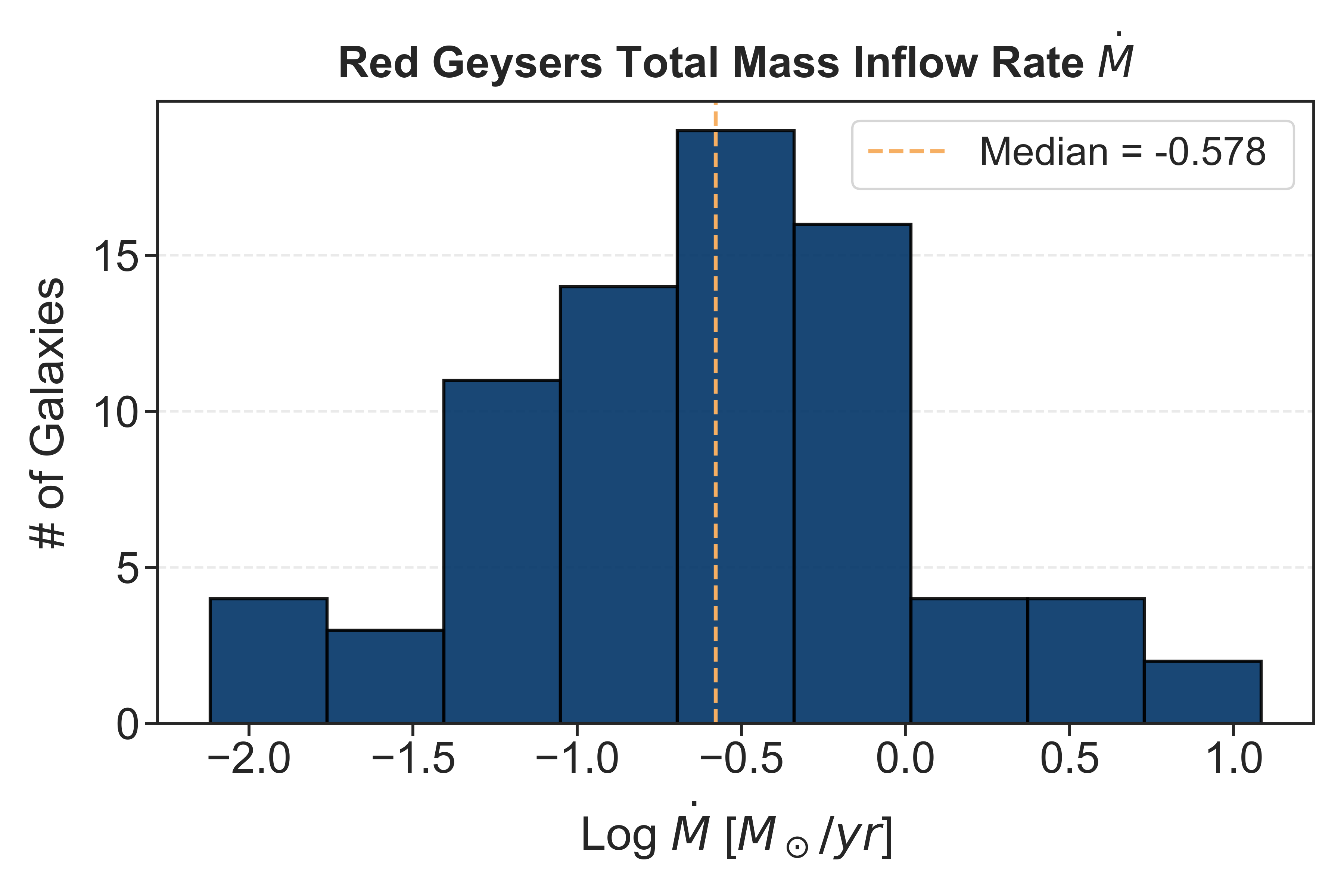} 
    \caption{Distribution of the total gas mass inflow rate $\dot M$ in each red geyser galaxy, including only galaxies with detected \NaID spaxels at a $S/N>5$ ($60\%$ of the sample) and removing the contributions from spaxels with outflowing velocities ($\sim 30\%$ of all detections). The distribution extends from $\sim 10^{-2} - 10^1M_\odot/yr$, and the median, shown in yellow, is found to be at $\log \dot M=-0.58$.
    }
    \label{fig:total_mass inflow rate}
\end{figure}

We carry out this calculation in all red geyser galaxies with at least one \NaID detection of $S/N>5$ ($\sim 60\%$ of the sample) and only use the spaxels with inflowing, positive velocities ($\sim 70\%$ of detections) to create a histogram of our entire sample, as shown in Figure \ref{fig:total_mass inflow rate}. The distribution has a median inflow rate of $\log \dot M\sim-0.6$ in $M_\odot/yr$ with $\dot M$ ranging from $\sim10^{-2}-10^1M_\odot/yr.$ The majority of the galaxies with detected inflowing spaxels are found to have mass inflow rates near the median ($\sim33\%$ within $\log\dot M\sim {-1}$ and $\log\dot M\sim {1}$), while less than $10\%$ of the sample has a mass inflow rate greater than $1M_\odot/yr$ or less than $\sim 0.1M_\odot/yr.$ This indicates that although large cool gas accretion is present in most red geysers, only a small fraction appear to be involved with strong accretion.

\begin{table*}
\centering
\setlength{\tabcolsep}{9pt}  

\begin{tabular}{lccccccccccc}
\hline
\hline
MaNGA ID & EW & $v_\mathrm{NaD}$ & $v_\mathrm{NaD}/v_\mathrm{ff}(3R_e)$ & $Cf$ &$\tau$ &$\log_{10} N_\mathrm{NaI}$ & $\log_{10} N_\mathrm{H}$ & $\dot{M}_\mathrm{in}$ & $M_\mathrm{in,tot}$ \\
 & [$\text{\AA}$] & [km s$^{-1}$] & & & & [cm$^{-2}$] & [cm$^{-2}$] & [$M_\odot$ yr$^{-1}$] & [$10^7 M_\odot$] \\
\hline
1-114230 & 0.72 & 16.07 & 0.05 & 0.10 & 2.61 & 13.63 & 21.16 & 0.05 & 0.18 \\
1-394355 & 0.89 & 68.05 & 0.09 & 0.08 & 3.69 & 14.14 & 21.31 & 0.79 & 1.04 \\
1-114245 & 1.39 & 25.83 & 0.12 & 0.15 & 2.13 & 13.91 & 21.24 & 0.93 & 1.85 \\
1-145922 & 1.24 & 73.51 & 0.14 & 0.16 & 2.12 & 13.78 & 21.21 & 2.30 & 2.62 \\
1-150792 & 0.88 & 82.93 & 0.14 & 0.09 & 4.60 & 14.41 & 21.39 & 0.23 & 0.40 \\
1-163594 & 0.88 & -6.47 & 0.05 & 0.11 & 1.14 & 13.69 & 21.18 & 0.09 & 0.30 \\
1-164007 & 0.92 & -88.75 & 0.14 & 0.08 & 1.74 & 14.07 & 21.29 & 0.07 & 0.03 \\
1-188530 & 0.95 & 48.43 & 0.07 & 0.14 & 1.33 & 13.61 & 21.16 & 0.85 & 1.45 \\
1-196372 & 0.89 & 14.77 & 0.07 & 0.09 & 3.02 & 14.06 & 21.29 & 0.10 & 0.27 \\
1-94168 & 2.16 & 38.23 & 0.09 & 0.31 & 1.64 & 13.63 & 21.16 & 2.20 & 4.68 \\
1-93908 & 1.23 & 7.28 & 0.02 & 0.12 & 1.40 & 13.89 & 21.24 & 0.15 & 1.33 \\
1-79747 & 1.09 & 21.27 & 0.06 & 0.13 & 0.92 & 13.58 & 21.15 & 0.31 & 0.71 \\
1-634825 & 1.65 & 62.54 & 0.11 & 0.22 & 1.52 & 13.69 & 21.18 & 2.51 & 3.37 \\
1-605863 & 0.39 & 86.41 & 0.17 & 0.04 & 3.21 & 14.06 & 21.29 & 0.14 & 0.08 \\
1-584723 & 1.42 & 14.79 & 0.12 & 0.15 & 1.90 & 13.86 & 21.23 & 0.65 & 0.85 \\
1-576738 & 0.85 & -0.97 & 0.07 & 0.08 & 2.02 & 14.01 & 21.27 & 0.04 & 0.13 \\
1-575742 & 1.65 & 16.62 & 0.03 & 0.21 & 1.16 & 13.67 & 21.17 & 0.54 & 2.97 \\
\hline
\end{tabular}

\caption{Measured \& calculated \NaID properties for a subset of red geyser galaxies with their MaNGA IDs. Columns list the equivalent width (EW), median \NaID velocity relative to host galaxy using double-Gaussian fitting ($v_{\rm NaD}$), median ratio of the observed velocity to the expected free-fall velocity assuming gases begin at $r_i=3R_e$, median \NaID covering fraction $C_f$, median \NaID optical depth $\tau$, median neutral sodium column density ($\log_{10}N_{\rm NaI}$), median neutral hydrogen column density ($\log_{10} N_{\rm H}$), total mass inflow rate ($\dot{M}_{\rm in}$) in regions with \NaID detections, and the total gas mass ($M_{\rm tot}$) in regions with \NaID detections.}
\label{table:calculations}
\end{table*}

We carry out these calculations for all red geyser galaxies, and a summary of our results is presented in Table \ref{table:calculations} for a subset of our sample. 

We note that the quantities we derive in this section are intended to provide a rough physical picture of the gases in red geyser galaxies, and, as described, their estimates rely on empirical conversions and several assumptions. Because these assumptions are associated with significant systemic uncertainties, the resulting values should only be interpreted as order-of-magnitude estimates rather than precise measurements.

\section{\textbf{Conclusion}}
\label{sec: Conclusion}
We study the cool neutral gas, traced by the \NaID absorption feature, in 140 quiescent galaxies known as red geysers with a median redshift of $z\sim0.04$. To isolate the interstellar component, we first remove the stellar absorption using the Data Analysis Pipeline (DAP), and then apply double-Gaussian fitting to extract the kinematics of the cool gases. From the ISM residual, we find that red geysers systematically host larger areas of cool, neutral gas reservoirs compared to control galaxies. More specifically, the detection fraction (requiring $\geq 5$ detected spaxels in each galaxy) in red geysers is $\sim63\%$ compared to the controls' $\sim 40\%$. Within galaxies with reliable \NaID detections ($\geq5$ detections with $S/N>5$), the area of \NaID reservoir in red geysers (median $4.96 \text{ kpc}^2$) is $\sim 1.6$ times larger than that of the control sample (median $3.06\text{ kpc}^2$).

Across our sample of red geyser galaxies, the \NaID absorption is predominantly ($\sim70\%$ of all detections) redshifted with a median inflow velocity of $\sim 47 \text{ km s}^{-1}$, and a median outflow velocity of $\sim -34 \text{ km s}^{-1}$. Additionally, these gas clouds are dynamically colder than the stellar component in their host galaxies, coherently flowing with a median $\frac{W50}{W50_*}\sim0.4$. 

We also show that the \NaID is centrally concentrated in red geysers, typically being found at an $R_{NaD}/R_e\sim 0.3.$ These cool, neutral gas clouds show no preferred orientation relative to the ionized bi-symmetric outflow features that define red geyser galaxies. 

Through comparing the kinematics of these cool gases with their host galaxy properties, we find that red geysers with stronger AGN activity (radio-detected) tend to host larger regions of cool gas: our analysis shows that regions of inflowing \NaID absorption on average become about 7 times larger when going from non-radio to radio-detected systems, while the blueshifted detections remain low in both systems, but their projected area still increases by a factor of $\sim6$.

Additionally, we show that red geysers with signs of interaction, morphological disturbances, or nearby neighbors host larger-scale inflow-classified \NaID compared to the non-interacting isolated system---on average having a $\sim2.7$ times larger area. Once again, the outflowing detections remain low across the different interaction classes, but still, the average area of the outflowing \NaID reservoirs increases by a factor of $\sim 4.8$ when comparing the clearly interacting galaxies to those that are isolated and undisturbed.


Our models show that the inflowing velocities are $\sim10\%$ of free-fall velocities of these gas clouds. Line-of-sight effect partially explains this discrepancy, but for the remainder, our estimates of terminal velocities suggests that drag from hot gases in the halo is likely not playing a major role in slowing down these cool gas clouds. 

The relatively low inflow velocities, the lack of \NaID absorption at large radii, and the fact that roughly two-thirds of red geysers show no evidence of interactions indicates that the cool neutral gas is not always supplied through external accretion. Our calculations for the timescale required to accelerate the gas clouds to observed velocities under gravity ($\sim 1$ Myr) and accretion timescales to reach the center and be disrupted ($\sim 20$ Myr) indicate that these \NaID clouds are short-lived and may originate from condensation and cooling of ionized gas within the central few kiloparsecs. Such an internal origin naturally explains both the limited spatial extent of the absorption and the modest velocities we observe.

We note a few limitations in our work. Our measurements of the kinematics of the cool, neutral gas relies on double-Gaussian fits to the \NaID doublet, which inevitably produces uncertainties, especially in the cases of more noisy spectra. We try to address and limit such scenarios by requiring the S/N for the detection to be greater than 5. Imposing a S/N threshold may limit our analysis to the central regions with more starlight and less noise, though in \S\,\ref{section:discussion} we show that while the S/N cuts increase the number of detected spaxels with Na I D, they do not lead to more extended detections in the outskirts of red geyser galaxies.

To quantify the uncertainties in our measurements, we use a Monte Carlo test and find that the median errors for our calculated kinematics are modest ($\pm13.84\text{ km s}^{-1}$ in velocity offsets and $7.86\%$ in dispersion); however, for individual spaxels, the errors can be larger, and degeneracies between the two Gaussian can affect both the velocity centroid and the linewidth.

In the context of these limitations, we find that interactions with nearby neighbors are an efficient mechanism for replenishing cool gas reservoirs in red geyser galaxies, feeding the central AGN, sustaining its activity, and regulating long-term quiescence.

This supports a bigger picture in which quiescent galaxies remain dormant through cycles of inflow, feedback, and regulation.

\begin{acknowledgments}
We thank the anonymous referee for their constructive feedback, which helped improve this manuscript.

A.M. gratefully acknowledges the support of the Rowland Summer Research Fellowship at Johns Hopkins University and the Vivien Thomas Scholars Initiative for providing funding and making the research possible during the summers of 2024 and 2025.

Funding for the Sloan Digital Sky 
Survey IV has been provided by the 
Alfred P. Sloan Foundation, the U.S. 
Department of Energy Office of 
Science, and the Participating 
Institutions. 

SDSS-IV acknowledges support and 
resources from the Center for High 
Performance Computing  at the 
University of Utah. The SDSS 
website is www.sdss4.org.

SDSS-IV is managed by the 
Astrophysical Research Consortium 
for the Participating Institutions 
of the SDSS Collaboration including 
the Brazilian Participation Group, 
the Carnegie Institution for Science, 
Carnegie Mellon University, Center for 
Astrophysics | Harvard \& 
Smithsonian, the Chilean Participation 
Group, the French Participation Group, 
Instituto de Astrof\'isica de 
Canarias, The Johns Hopkins University, Kavli Institute for the 
Physics and Mathematics of the 
Universe (IPMU) / University of 
Tokyo, the Korean Participation Group, 
Lawrence Berkeley National Laboratory, 
Leibniz Institut f\"ur Astrophysik 
Potsdam (AIP),  Max-Planck-Institut 
f\"ur Astronomie (MPIA Heidelberg), 
Max-Planck-Institut f\"ur 
Astrophysik (MPA Garching), 
Max-Planck-Institut f\"ur 
Extraterrestrische Physik (MPE), 
National Astronomical Observatories of 
China, New Mexico State University, 
New York University, University of 
Notre Dame, Observat\'ario 
Nacional / MCTI, The Ohio State 
University, Pennsylvania State 
University, Shanghai 
Astronomical Observatory, United 
Kingdom Participation Group, 
Universidad Nacional Aut\'onoma 
de M\'exico, University of Arizona, 
University of Colorado Boulder, 
University of Oxford, University of 
Portsmouth, University of Utah, 
University of Virginia, University 
of Washington, University of 
Wisconsin, Vanderbilt University, 
and Yale University.

The Legacy Surveys consist of three individual and complementary projects: the Dark Energy Camera Legacy Survey (DECaLS; Proposal ID \#2014B-0404; PIs: David Schlegel and Arjun Dey), the Beijing-Arizona Sky Survey (BASS; NOAO Prop. ID \#2015A-0801; PIs: Zhou Xu and Xiaohui Fan), and the Mayall z-band Legacy Survey (MzLS; Prop. ID \#2016A-0453; PI: Arjun Dey). DECaLS, BASS and MzLS together include data obtained, respectively, at the Blanco telescope, Cerro Tololo Inter-American Observatory, NSF’s NOIRLab; the Bok telescope, Steward Observatory, University of Arizona; and the Mayall telescope, Kitt Peak National Observatory, NOIRLab. Pipeline processing and analyses of the data were supported by NOIRLab and the Lawrence Berkeley National Laboratory (LBNL). The Legacy Surveys project is honored to be permitted to conduct astronomical research on Iolkam Du’ag (Kitt Peak), a mountain with particular significance to the Tohono O’odham Nation.

NOIRLab is operated by the Association of Universities for Research in Astronomy (AURA) under a cooperative agreement with the National Science Foundation. LBNL is managed by the Regents of the University of California under contract to the U.S. Department of Energy.

This work also uses data obtained with the Dark Energy Camera (DECam), which was constructed by the Dark Energy Survey (DES) collaboration. Funding for the DES Projects has been provided by the U.S. Department of Energy, the U.S. National Science Foundation, the Ministry of Science and Education of Spain, the Science and Technology Facilities Council of the United Kingdom, the Higher Education Funding Council for England, the National Center for Supercomputing Applications at the University of Illinois at Urbana-Champaign, the Kavli Institute of Cosmological Physics at the University of Chicago, Center for Cosmology and Astro-Particle Physics at the Ohio State University, the Mitchell Institute for Fundamental Physics and Astronomy at Texas A\&M University, Financiadora de Estudos e Projetos, Fundacao Carlos Chagas Filho de Amparo, Financiadora de Estudos e Projetos, Fundacao Carlos Chagas Filho de Amparo a Pesquisa do Estado do Rio de Janeiro, Conselho Nacional de Desenvolvimento Cientifico e Tecnologico and the Ministerio da Ciencia, Tecnologia e Inovacao, the Deutsche Forschungsgemeinschaft and the Collaborating Institutions in the Dark Energy Survey. The Collaborating Institutions are Argonne National Laboratory, the University of California at Santa Cruz, the University of Cambridge, Centro de Investigaciones Energeticas, Medioambientales y Tecnologicas-Madrid, the University of Chicago, University College London, the DES-Brazil Consortium, the University of Edinburgh, the Eidgenossische Technische Hochschule (ETH) Zurich, Fermi National Accelerator Laboratory, the University of Illinois at Urbana-Champaign, the Institut de Ciencies de l’Espai (IEEC/CSIC), the Institut de Fisica d’Altes Energies, Lawrence Berkeley National Laboratory, the Ludwig Maximilians Universitat Munchen and the associated Excellence Cluster Universe, the University of Michigan, NSF’s NOIRLab, the University of Nottingham, the Ohio State University, the University of Pennsylvania, the University of Portsmouth, SLAC National Accelerator Laboratory, Stanford University, the University of Sussex, and Texas A\&M University.

BASS is a key project of the Telescope Access Program (TAP), which has been funded by the National Astronomical Observatories of China, the Chinese Academy of Sciences (the Strategic Priority Research Program “The Emergence of Cosmological Structures” Grant \# XDB09000000), and the Special Fund for Astronomy from the Ministry of Finance. The BASS is also supported by the External Cooperation Program of Chinese Academy of Sciences (Grant \# 114A11KYSB20160057), and Chinese National Natural Science Foundation (Grant \# 12120101003, \# 11433005).

The Legacy Survey team makes use of data products from the Near-Earth Object Wide-field Infrared Survey Explorer (NEOWISE), which is a project of the Jet Propulsion Laboratory/California Institute of Technology. NEOWISE is funded by the National Aeronautics and Space Administration.

The Legacy Surveys imaging of the DESI footprint is supported by the Director, Office of Science, Office of High Energy Physics of the U.S. Department of Energy under Contract No. DE-AC02-05CH1123, by the National Energy Research Scientific Computing Center, a DOE Office of Science User Facility under the same contract; and by the U.S. National Science Foundation, Division of Astronomical Sciences under Contract No. AST-0950945 to NOAO.

\end{acknowledgments}

\bibliography{citations}{}
\bibliographystyle{aasjournalv7}



\end{document}